\documentclass[sigconf, nonacm]{acmart}

\newcommand\vldbdoi{10.14778/3425879.3425891}
\pdfoutput=1
\newcommand\vldbvolume{14}
\newcommand\vldbissue{2}
\newcommand\vldbyear{2021}
\newcommand\vldbauthors{\authors}
\newcommand\vldbtitle{\shorttitle} 

\newcommand\vldbpagestyle{empty}
\newcommand\vldbpages{215-227}
\usepackage[T1]{fontenc}
\usepackage{graphicx}
\usepackage{amsthm}
\usepackage{dcolumn}
\usepackage{subfig}
\usepackage{multirow}
\usepackage{algorithm}
\usepackage[noend]{algorithmic}
\usepackage{pgfplots}
\usepackage{filecontents}
\usepackage{amsmath}
\usepackage{amsthm}
\usepackage[a-1b]{pdfx}   
\usepackage{balance}

\usepackage{hyperref}
\newcommand{\RNum}[1]{\uppercase\expandafter{\romannumeral #1\relax}}

\newtheorem{innercustomgeneric}{\customgenericname}
\providecommand{\customgenericname}{}
\newcommand{\newcustomtheorem}[2]{%
  \newenvironment{#1}[1]
  {%
  \renewcommand\customgenericname{#2}%
  \renewcommand\theinnercustomgeneric{##1}%
  \innercustomgeneric
  }

  {\endinnercustomgeneric}
}
\newcustomtheorem{customlemma}{Lemma}

\begin{document}
\title{PPQ-Trajectory: Spatio-temporal Quantization for Querying in Large Trajectory Repositories}

\author{Shuang Wang}
\affiliation{%
  \institution{University of Warwick}
  \city{Coventry}
  \country{United Kingdom}
}
\email{Shuang.Wang.1@warwick.ac.uk}

\author{Hakan Ferhatosmanoglu}
\affiliation{%
  \institution{University of Warwick}
  \city{Coventry}
  \country{United Kingdom}
}
\email{Hakan.F@warwick.ac.uk}

\begin{abstract}
We present PPQ-trajectory, a spatio-temporal quantization based solution for querying large dynamic trajectory data. PPQ-trajectory includes a partition-wise predictive quantizer (PPQ) 
that generates an error-bounded codebook with autocorrelation and spatial proximity-based partitions. The codebook is indexed to run approximate and exact spatio-temporal queries over compressed trajectories.  PPQ-trajectory  includes a coordinate quadtree coding for the codebook with support for exact queries.
An incremental temporal partition-based index is utilised to avoid full reconstruction of trajectories during queries. An extensive set of experimental results for spatio-temporal queries on real trajectory datasets is presented. PPQ-trajectory shows significant improvements over the alternatives with respect to several performance measures, including the accuracy of results when the summary is used directly to provide approximate query results, the spatial deviation with which spatio-temporal path queries can be answered when the summary is used as an index,
and the time taken to construct the summary. Superior results on the quality of the summary and the compression ratio are also demonstrated.
\end{abstract}
\maketitle

\pagestyle{\vldbpagestyle}
\begingroup\small\noindent\raggedright\textbf{PVLDB Reference Format:}\\
\vldbauthors. \vldbtitle. PVLDB, \vldbvolume(\vldbissue): \vldbpages, \vldbyear.\\
\href{https://doi.org/\vldbdoi}{doi:\vldbdoi}
\endgroup
\begingroup
\renewcommand\thefootnote{}\footnote{\noindent
This work is licensed under the Creative Commons BY-NC-ND 4.0 International License. Visit \url{https://creativecommons.org/licenses/by-nc-nd/4.0/} to view a copy of this license. For any use beyond those covered by this license, obtain permission by emailing \href{mailto:info@vldb.org}{info@vldb.org}. Copyright is held by the owner/author(s). Publication rights licensed to the VLDB Endowment. \\
\raggedright Proceedings of the VLDB Endowment, Vol. \vldbvolume, No. \vldbissue\ %
ISSN 2150-8097. \\
\href{https://doi.org/\vldbdoi}{doi:\vldbdoi} \\
}\addtocounter{footnote}{-1}\endgroup

\section{Introduction}
With the prevalence of positioning devices and mobile services, massive amounts of location sequences are being generated continuously. 
Maintaining and querying small-sized representations of raw trajectory data are needed for a wide variety of applications, such as real-time traffic management \cite{ning2019vehicular} and intelligent transport systems \cite{cai2017vector}.

Existing trajectory compression methods do not address this need for a number of reasons. First, many of them are defined for edge sequences in a road network  \cite{funke2019pathfinder,han2017compress, koide2018enhanced,song2014press}. They require pre-processing steps of mapping raw GPS data to the road network structure, followed by transforming the map-matched location data to edge-based sequences. The mapping and transformation processes reduce accuracy and result in limited support for detailed queries. Second, most solutions perform offline compression over full trajectory data, with execution times usually undesirable for online applications. There is a need for scalable online compression. Third, the existing compressed representations can not be directly used to answer spatio-temporal queries without a costly decompression process. 

\begin{figure}[b]
  \centering
  \hspace{4mm}
  \includegraphics[height=4.1cm,width=8.5cm]{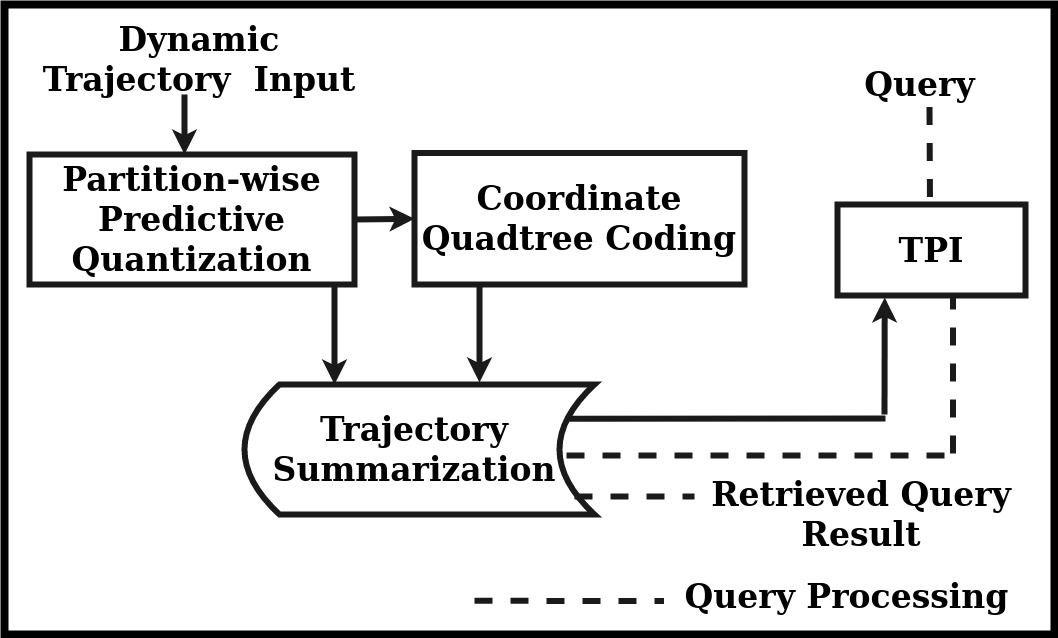}
  \caption{An Overview of PPQ-trajectory}
  \label{overview}
\end{figure}

To address these challenges, we present PPQ-trajectory, a spatio-temporal quantization-based solution to generate a compact representation and support a wide range of queries over large trajectory data. 
An overview of PPQ-trajectory is presented in Figure \ref{overview}.

The first part of PPQ-trajectory is the partition-wise predictive quantizer (PPQ) that generates an error-bounded summary, consisting of the codebook and prediction coefficients for spatial and autocorrelation-based partitions. The second part is the coordinate quadtree coding (CQC) for the error space caused by the quantization, which enables an accurate reconstruction of the trajectories. These two parts form the summary for the trajectory data, as illustrated in Figure \ref{overview}. The third part is the temporal partition-based index (TPI) for organizing the quantized spatio-temporal data. Given a query, TPI is used to prune the data space and generate a candidate list of trajectories, whose reconstructed points can be computed from the summary.
Overall, PPQ-trajectory generates and uses an indexed summary over raw data sequences to support efficient analysis, ranging from simple queries, such as vehicles passing by a location $(x,y)$ at a given time $t$, to more complex analytic tasks, such as predicting future positions of entities.
 
We evaluate PPQ-trajectory with respect to a number of performance measures: the quality of approximate query results, efficiency of exact queries, index building times, and compression ratio. We implemented several baselines, including the widely-used product quantization~\cite{jegou2010product}, residual quantization~\cite{chen2010approximate}, REST \cite{zhao2018rest}, which is a recent reference based trajectory compression method,  and TrajStore \cite{cudre2010trajstore}, an adaptive storage solution for trajectories. 

This paper makes the following contributions: 
(1) A spatio-temporal predictive quantizer, PPQ, is designed with an error-bounded codebook for each of the partitions, which are incrementally generated based on spatial and autocorrelation similarity.
(2) Utilizing a quadtree structure and a padding process, CQC is developed to encode the relative positions of trajectory points with reconstructed ones for an accurate trajectory reconstruction. A local search strategy is presented to identify exact query results.
(3) The temporal partition-based index dynamically reuses parts of the past index to support efficient spatio-temporal queries.
(4) PPQ-trajectory answers spatio-temporal queries over raw data, without a full reconstruction of trajectories or accessing all of the candidate trajectories. 
(5) Experimental results demonstrate significant improvements achieved by PPQ-trajectory. For example, for approximate spatio-temporal queries, it is $3\%-52\%$ more accurate, 
compared to product quantization, residual quantization, and TrajStore.
The  mean absolute error (MAE) 
of PPQ-trajectory is a few or tens of meters, while the alternative approaches' MAE
values are orders of magnitude larger for the same size codebook. Significant improvements are also observed for the efficiency, index building times, and compression ratios.

The rest of the paper is organized as follows. The related work is in Section \ref{related_work}.  Section \ref{online_summarization} presents PPQ. In Section \ref{optimal_solution}, we present CQC and the associated local search strategy. In Section \ref{TSDI}, the temporal organization for the quantized data is presented. The effectiveness of PPQ-trajectory is verified via an extensive set of experiments in Section \ref{experiments}. Conclusion is provided in Section \ref{conclusion}.

\vspace{-2mm}
\section{Related Work}
\label{related_work}
With the widespread adoption of location-based services, compressing trajectory data has become a prevalent area with high practical relevance \cite{bellman2015applied,koide2018enhanced,liu2016novel}. While traditional compression methods aim to reduce the reconstruction error and improve compression ratio, the data management challenge is to design the compression method with the objective of answering queries efficiently, and supporting online querying directly over compressed data.

Road network-constrained trajectory compression has gained significant attention \cite{funke2019pathfinder,han2017compress,koide2018enhanced,liu2015bounded,popa2015spatio,song2014press}. The common approach is to map raw trajectories to road networks and compress the map-matched trajectories \cite{kellaris2013map}. 
There is also significant attention on raw trajectory compression. SQUISH and SQUISH-E use a priority queue to remove redundant points  \cite{muckell2011squish,muckell2014compression}. A bounded quadrant system (BQS) is developed in \cite{liu2015bounded}, which uses convex hull bounding to achieve trajectory compression. Based on BQS, \cite{liu2016novel} achieves streaming trajectory compression, and aging history data without overwriting. Another recent method transforms trajectories into vehicle state vector functions and generates an inverted index based on the road segments \cite{cai2017vector}. This approach requires road segment information and matching each road id with corresponding objects.

The solutions that we included in our experiments are TrajStore \cite{cudre2010trajstore} and REST \cite{zhao2018rest}. TrajStore aims compression via an adaptive spatial index and clustering the sub-trajectories. It recursively updates the index by merging, splitting or appending. 
REST is a recent compression-based method which compares trajectories with the sub-trajectories of a reference set. Generating a representative set is challenging especially under changing conditions, where it can fail to represent data from regions that lack enough samples.

Early work in this area applies traditional index structures for trajectory data. For example, STRIPES uses quadtrees to index the predicted positions of moving objects \cite{patel2004stripes}. In \cite{cai2004indexing}, an index for trajectories is developed by indexing the coefficients of Chebyshev polynomials that represent trajectories. Most those methods focus on similarity queries and do not address efficient spatio-temporal database queries. Zheng et al., \cite{zheng2019reference} index the reference-based trajectories with IR-tree \cite{li2010ir}, which is based on R-tree referring to the inverted files for sub-trajectories.  

Quantization is a popular method for traditional compression, and for nearest neighbor searches on multi dimensional data, especially for multimedia and computer vision applications \cite{ferhatosmanoglu2000vector,liu2003efficient,norouzi2013cartesian,tuncel2002vq,weber1998quantitative}.  Predictive quantization has been applied for online summarization of multiple one-dimensional data streams \cite{altiparmak2007incremental}. The correlation among consecutive points is employed to predict current points, then the prediction errors are summarized into a smaller number of bits \cite{altiparmak2007incremental}.  Product Quantization and Residual Quantization \cite{chen2010approximate,ge2013optimized,jegou2010product} have made significant impact on approximate nearest neighbor searching in computer vision applications. We included these two methods in our performance evaluation. There have been some work to use quantization for encoding trajectories \cite{chen2012compression}, transforming differential trajectory points into strings for compression \cite{lv2015trajectory}, and 
retaining information for trajectory prediction \cite{chan2012utilizing}. These methods adopt compression but with no particular support for efficient querying over compressed trajectories.  
Our goal is to quantize dynamic trajectories into an error-bounded and query friendly representation, where there is neither need to fully reconstruct nor traverse the full trajectories. Trajectory data is summarized online, exploiting their large-scale nature, for the purpose of efficient query processing.

\section{Online Quantization in \\PPQ-Trajectory}
\label{online_summarization}


In this section, we present our spatio-temporal quantization based summarization process.
The performance measures are  the accuracy of results when the summary is used directly to provide approximate query results, the spatial deviation with which spatio-temporal path queries can be answered when the summary is used as an index,  and the time taken to construct the summary. The quality of the summary and the compression ratio are also related measures.  
Table \ref{summary_of_notations} summarizes the notation used throughout the paper. Basic definitions of trajectories and codebooks are as follows.
\begin{definition}
(Trajectory) A trajectory $T$ is a finite sequence of time-stamped positions in the form of (($x_1$, $y_1$, $t_1$), ($x_2$, $y_2$, $t_2$), ..., ($x_n$, $y_n$, $t_n$)), where $0 \leq t_i \leq t_j \leq t_n$ with $ 0\leq i \leq j \leq n$.
\end{definition}

\begin{definition}
(Error-bounded Codebook)
\label{Bounded-Codebook}
Consider a codebook $C$ = $\{C_1,...,C_n\}$ where the set of trajectory points $\mathcal{T}^i$ is indexed by codeword $C_i$. For any trajectory point $T_j^t \in \mathcal{T}^i$, if $\left\|T_j^t-C_i\right\|_2$ $\leq$ $\varepsilon_1$, then the codebook $C$ is bounded with $\varepsilon_1$. 
\end{definition}

\begin{table}
    \centering
    \caption{Summary of Notation}
    \label{summary_of_notations}
\setlength{\tabcolsep}{1mm}{
    \begin{tabular}{ll}
        	\hline
        \textbf{Variable} & \textbf{Definition}\\
       	\hline
       $T_i$ & the $i$-th trajectory\\ 
       $T_i^t$ & trajectory point of $T_i$ at time $t$
        \\
        $T^t$ & trajectory points at time $t$
        \\
       $\widehat{T}_i^t$ & reconstructed trajectory point of $T_i^t$\\
       $\widetilde{T}_{i}^t$ & predicted trajectory point of $T_i^t$\\
       $e_i^t$ & \multirow{1}{7cm}{prediction error of $T_i^t$, i.e., $T_i^t - \tilde{T}_{i}$}
       \\
       $f$ & a prediction function\\
\bottomrule 
    \end{tabular}}
\end{table}
\subsection{Error-bounded Predictive Quantization}
\label{minimize_information_loss}
We first present the \underline{e}rror bounded \underline{p}redictive \underline{q}uantizer
(E-PQ) to obtain a compact codebook for trajectories.
Predictive quantization (PQ) has been successfully applied for one-dimensional data streams by quantizing the error of the estimate of the sample at time $t$ with previous $k$ samples, i.e., $\widetilde{x}[t] = f(x[t-1], x[t-2],...,x[t-k])$ \cite{altiparmak2007incremental,fletcher2007robust}. A prediction function $f$ is learned over training data, and a codebook is generated via a vector quantizer \cite{tuncel2002vq} on the prediction errors by assigning them to the nearest centroids of their clusters. The range of the error $e[t] = x[t] - \widetilde{x}[t]$ is narrower than the original data which enables the errors to be quantized more effectively than the original data \cite{altiparmak2007incremental}. 

To estimate the trajectory points using their correlations, we define a prediction function as an extension to the case for one-dimensional streams \cite{altiparmak2007incremental}. For ease of demonstration, we define $f$ as a linear  model that predicts ${T}^{t}_i$ using the previous $k$ samples.  The prediction is computed as:
\begin{equation}
\begin{aligned}
& \underset{f}{\text{min}}
\sum_{i=1}^N  \left\|{T}^{t}_i-f({T}^{(t-k:t-1)}_i)\right\|_2\label{eq:r1}
\end{aligned}
\end{equation}
where $T^{t}_i$ represents the position $(x_t, y_t) $ of $T_i$ at time $t$, ${T}^{(t-k:t-1)}_i$ is the sequence of the trajectory $T_i$ at time interval $[t-k,t-1]$, and $f$ denotes the prediction model.

The prediction error $e^t_i$ is defined as: 
\begin{equation}
\begin{aligned}
& e^t_i = T^{t}_i-\widetilde{T}^{t}_i  \label{eq:predicterror},
&\widetilde{T}^{t}_i = \sum_{j=1}^kP_{j}[t]\widehat{T}^{{t-j}}_i
\end{aligned}
\end{equation}
where $\widetilde{T}^{t}_i $ is the prediction of $T^{t}_i$, $P_{j}[t]$ is the $j$-th prediction coefficient of $f$, and $\widehat{T}^{{t-k}}_i$ is the reconstruction of ${T}^{{t-k}}_i$.

The prediction errors $\{e^t_i\}$ can be summarized into an error-bounded codebook $C$ as: 

\begin{equation}
\begin{aligned}
&\quad \quad \quad \quad\text{min}  |C|\\
 \label{eq:r3}
 &\text{s.t.} {\left\|e^{t}_i-C(b^{t}_i)\right\|_2} \leq \varepsilon_1, 
b^{t}_i \in \{1,..., V\}
 \end{aligned}
 \end{equation}
where $C=\{C_1, C_2,...,C_V \}$ is the error-bounded codebook, $V$ denotes the size of $C$,
$b^{t}_i$ is the codeword index for $e_i^t$, and $C(b^{t}_i)$ denotes the codeword assigned to represent  $e^{t}_i$. 
Every codeword $C_i \in C$ is a cluster centroid, obtained by partitioning the data space to facilitate indexing and compression.
Equation \ref{eq:r3} aims to achieve a minimal error-bounded codebook $C$ for the given $\varepsilon_1$, which is non-convex. For dynamic databases, $C$ needs to be incrementally updated with evolving $t$ values. In order to get the approximate solution, at time $t+1$, if part of the prediction errors $\{e^{t+1}_i\}$ can not satisfy the threshold, the additional codewords are added to update $C$ to guarantee the boundary requirement continuously.

The reconstructed $T^{t}_i$, $ \hat{T}^t_i$, is obtained where: 
 \begin{equation}
\begin{aligned}
& \widehat{T}^t_i = \widetilde{T}^{t}_i + C(b^{t}_i) \label{eq:r5} 
\end{aligned}
\end{equation}

The procedure of quantizing dynamic trajectories is summarized in Algorithm \ref{predictive_spatio_temporal_quantization}.  In Line 3, the  prediction coefficient $P_j[t]$ can be solved in a standard manner \cite{altiparmak2007incremental, gersho2012vector}.
Line 4 denotes the prediction of the $t$-th trajectory point by its previous $k$ reconstructed points. For the time $t\leq k$, $P_j[t]$ is set to zero.  
At Line 6, $Incremental\_Quantizer$ represents the quantization process of Equation \ref{eq:r3}. E-PQ maps the trajectory data into $\{P_j[t]\}$, $C$, $\{b_i^t\}$.


\begin{algorithm}[tb]
	\renewcommand{\algorithmicrequire}{\textbf{Input:}}
	\renewcommand{\algorithmicensure}{\textbf{Output:}}
	\caption{Error-bounded Predictive Quantization}
	\label{predictive_spatio_temporal_quantization}
	\begin{algorithmic}[1]
		\REQUIRE $\{T^{t}_i\}$, $\varepsilon_1$
		\ENSURE $\{P_j[t]\}$, $C$, $\{b_i^t\}$
		\STATE t = 1 
		\WHILE{$\{T^{t}_i\}$ is not empty}
		 	\STATE  Derive $\{P_j[t]\}$ based on Equation \ref{eq:r1}
		 	\STATE $\widetilde{T}^{t}_i = \sum_{j=1}^kP_{j}[t]\widehat{T}^{{t-j}}_i$
		 	\STATE $\{e_i^t\} = \{T^{t}_i\}-\{\widetilde{T}^{t}_i$\}
		 	\STATE  $C$, $\{b_i^t\}$  = $Incremental\_Quantizer$($\{e_i^t\}$, $C$,  $\varepsilon_1$)
		 	\STATE  $\hat{T}^t_i$ = $\widetilde{T}^{t}_i$ + $C(b^{t}_i)$
            \STATE t = t + 1
	\ENDWHILE
	\end{algorithmic}  
\end{algorithm}

\subsection{Partition-wise Predictive Quantization}
\label{minimize_bounded_codebook}
We now present our quantizer that partitions trajectory points and applies E-PQ for each partition. The partition-wise predictive quantization (PPQ) is formulated as:
\begin{equation}
 \begin{aligned}
 &\mathcal{N}^t = \{ \mathcal{N}^t_1,...,\mathcal{N}^t_q\}
 \label{eq:partition}
 \end{aligned}
 \end{equation}
 \begin{equation}
 \begin{aligned}
 &\text{min}\sum_{i\in \mathcal{N}^t_j} \left\|{T}^{t}_i-f_j(T^{t-k:t-1}_i)\right\|_2, \mathcal{N}^t_j \in \mathcal{N}^t
 \label{eq:partition_quantization}
 \end{aligned}
\end{equation} 
In Equation \ref{eq:partition}, $N$ trajectory points $\{T^t_i\}$ are partitioned into $q$ subsets, where $\mathcal{N}^t_j$ denotes the set of trajectory IDs assigned to the $j$-th partition. 
$f_j \in \{f_1,...f_q\}$ is the prediction function for $\mathcal{N}^t_j$.
This enables the use of a single prediction function $f_j$ for trajectory points of $\mathcal{N}^t_j$,
and is resolved by Equation \ref{eq:partition_quantization}.
Via $\{f_1,...f_q\}$, the correlation among consecutive trajectory points in every $\mathcal{N}^t_j \in \mathcal{N}^t$ is modeled by a specific $f_j$,  then the dynamic range of prediction errors is further narrowed down. When $q$ = 1, Equation \ref{eq:r1} and  Equation \ref{eq:partition_quantization} become the same. Similarly for the $\{e^t_i\}$ obtained from multiple predictions, they are summarized with Equation \ref{eq:r3}.

\subsubsection{Partitioning for Grouped Modeling}
\label{partition_for_group_model}

We partition the trajectory points using their spatial and autocorrelation similarities. Tobler's first law of geography indicates ``everything is related to everything else, but nearby things are more related than distant things" \cite{tobler1979cellular}. Hence, assigning trajectory points based on spatial proximity is a natural approach to be able to use the same $f_j$  to model them. However, as the role of $f_j$ is to capture the correlations between consecutive trajectory points, 
assigning trajectory points with similar autocorrelations to $\mathcal{N}^t_j$ can enable a more accurate prediction by $f_j$. In our setting, the correlation between $T_i^t$ and $T_i^{t-k:t-1}$ follows an autoregressive process of order $k$ (AR($k$)) \cite{cryer1991time, papoulis2002probability}, where the current trajectory point ($T_i^t$) is linearly related to the lagged $k$ points ($T_i^{t-k:t-1}$). We derive the parameters of AR($k$) as $\{a^t_i\}$ and utilize them to quantity the lag-$k$ autocorrelation. Assigning trajectory points with similar $\{a^t_i\}$ to the same partition $\mathcal{N}^t_j$ allows $f_j$ to more effectively capture the correlations between consecutive trajectory points. 

The partitioning process is repeated until all partitions satisfy Equations \ref{eq:ppq_spatial_partition} and \ref{eq:ppq_autocorrelation_partition}, for spatial and autocorrelation similarity, respectively. For the spatial proximity,  the deviation between any point in $\mathcal{N}_j$ and the centroid of $\mathcal{N}_j$ should be less than $\varepsilon_p$, otherwise,  $q$ increases until Equation \ref{eq:ppq_spatial_partition} is satisfied. 

\begin{equation}
 \begin{aligned}
\left\|T^{t}_i-centroid_S(\mathcal{N}^t_j)\right\|_2 \leq \varepsilon_p, \text{for all } \mathcal{N}^t_j \in \mathcal{N}^t, i \in \mathcal{N}^t_j
 \label{eq:ppq_spatial_partition}
 \end{aligned}
\end{equation}

 
Similarly, for the autocorrelation similarity, the 
partitions satisfy Equation \ref{eq:ppq_autocorrelation_partition}, where $a^{t}_i$ represents the lag-$k$ autocorrelation of $T^t_i$, and $centroid_A(\mathcal{N}^t_j)$ is the centroid of the autocorrelation of trajectory points in $\mathcal{N}_j$. 

\begin{equation}
 \begin{aligned}
\left\|a^{t}_i-centroid_A(\mathcal{N}^t_j)\right\|_2 \leq \varepsilon_p,  \text{for all } \mathcal{N}^t_j \in \mathcal{N}^t, i \in \mathcal{N}^t_j
 \label{eq:ppq_autocorrelation_partition}
 \end{aligned}
\end{equation}
The setting of $\varepsilon_p$ is based on the size of the region that trajectories span (for spatial proximity), or the magnitude and distribution of autocorrelation coefficients (for autocorrelation similarity).

The computational complexity for $q$ partitions is $\mathcal{O}(qmNl)$ as shown 
in Lemma \ref{complexity_partitioning_PQ}, where $N$ is the number of trajectory points, $m$ denotes the rounds of increasing $q$ to satisfy Equation \ref{eq:ppq_spatial_partition} or \ref{eq:ppq_autocorrelation_partition}, and $l$ represents the number of iterations to obtain the given number of partitions by K-means \cite{lloyd1982least}. The complexity is proportional to $q$.

\subsubsection{Incremental Temporal Partitioning} 
\label{incremental_temporal_partitioning}
Consider the partitions at time $t$, i.e., $\mathcal{N}^t = \{ \mathcal{N}^t_1,...,\mathcal{N}^t_q\}$. Instead of performing partitioning from scratch, an incremental partitioning for time $t+1$ is performed with the following steps. First, every trajectory point at time $t+1$, i.e., $\{T^{t+1}_i\}$, is assigned to the same partition as $T^t_i$. Second, when a partition, $\mathcal{N}^{t+1}_j$, does not satisfy the requirement for $\varepsilon_p$,
a new partitioning is performed over trajectory points in $\mathcal{N}^{t+1}_j$ until the resultant partition satisfies the requirement. 
Third, with ${\mathcal{N}^{t+1}}_j$ and ${\mathcal{N}^{t+1}}_{j'}$, if $\left\|centroid_{S/A}({\mathcal{N}^{t+1}}_j) - centroid_{S/A}(\mathcal{N}^{t+1}_{j'})\right\|_2 \leq \varepsilon_p$, we merge ${\mathcal{N}^{t+1}}_j'$ to ${\mathcal{N}^{t+1}}_{j}$ to avoid too many fragmented partitions.
Specifically, for every partition $\mathcal{N}^{t+1}_j$, we only allow merging at most once, as excessive merging might influence the preciseness of partitioning and the quantization performance. 
 If there are $N'$ trajectory points at $t+1$ that do not satisfy the requirement for $\varepsilon_p$,  $q'$ new partitions are generated via an $m'$ rounds of checking with Equation \ref{eq:ppq_spatial_partition} or \ref{eq:ppq_autocorrelation_partition}, then
 the computational complexity of the incremental temporal partitioning is $\mathcal{O}(q'm'N'l + q'q)$, which is presented in LEMMA \ref{incremental_complexity_partitioning_PQ}. $q'$ is only relevant to the distribution of the $N'$ trajectory points.
$N'$ gets smaller when the points among the consecutive timestamps are highly similar in autocorrelations or spatially close. In the worst case, when all the $N$ trajectory points at time $t+1$ do not satisfy the $q$ partitions at time $t$, i.e., $N$ = $N'$, the computational complexity of incremental temporal partitioning becomes $\mathcal{O}(qmNl)$.
\begin{customlemma}{1}\label{complexity_partitioning_PQ}
    The computational complexity of partitioning $\{T^{t}_i\}$ into $q$ partitions (Section \ref{partition_for_group_model}) is $\mathcal{O}(qmNl)$.
    \end{customlemma} 
 \begin{proof}
Let $q_i$ ($i \in [1, m]$) be the number of partitions at the $i$-th round, which increases by $a$ at every round until all the partitions satisfy Equation \ref{eq:ppq_spatial_partition} or \ref{eq:ppq_autocorrelation_partition}.  Hence, $q$ = $q_m$ = $ma$. 
 For the $i$-th round, the computational complexity is the same as partitioning $N$ trajectory points into $q_i$ clusters, i.e., $\mathcal{O}(q_iNl)$.  Then, the overall computational cost is $aNl$ + ... + $maNl$ = $\frac{am(1+m)}{2}Nl$, i.e., $\mathcal{O}(qmNl)$.
 \end{proof}
  \begin{customlemma}{2}\label{incremental_complexity_partitioning_PQ}
    The computational complexity of incremental temporal partitioning for $\{T^{t+1}_i\}$ is $\mathcal{O}(q'm'N'l +  q'q)$.
    \end{customlemma} 
 \begin{proof}
 According to LEMMA \ref{complexity_partitioning_PQ}, the complexity of partitioning $N'$ trajectory points into $q'$ partitions is  $\mathcal{O}(q'm'N'l)$. For $q'$ new partitions at time $t+1$, in the worst case, there will be $q$ + ... + $(q-(q'-1))$ = $\frac{q'(q+(q-(q'-1)))}{2}$ computations to check if a new partition can be merged into the existing $q$ partitions. Hence, the overall complexity is $\mathcal{O}(q'm'N'l + q'q)$.  
 \end{proof}

\section{Local Coding in Error-bounded Codebook}
\label{optimal_solution}
 
\begin{figure}[tb]
\centering
  \includegraphics[height=3.5cm,width=7.5cm]{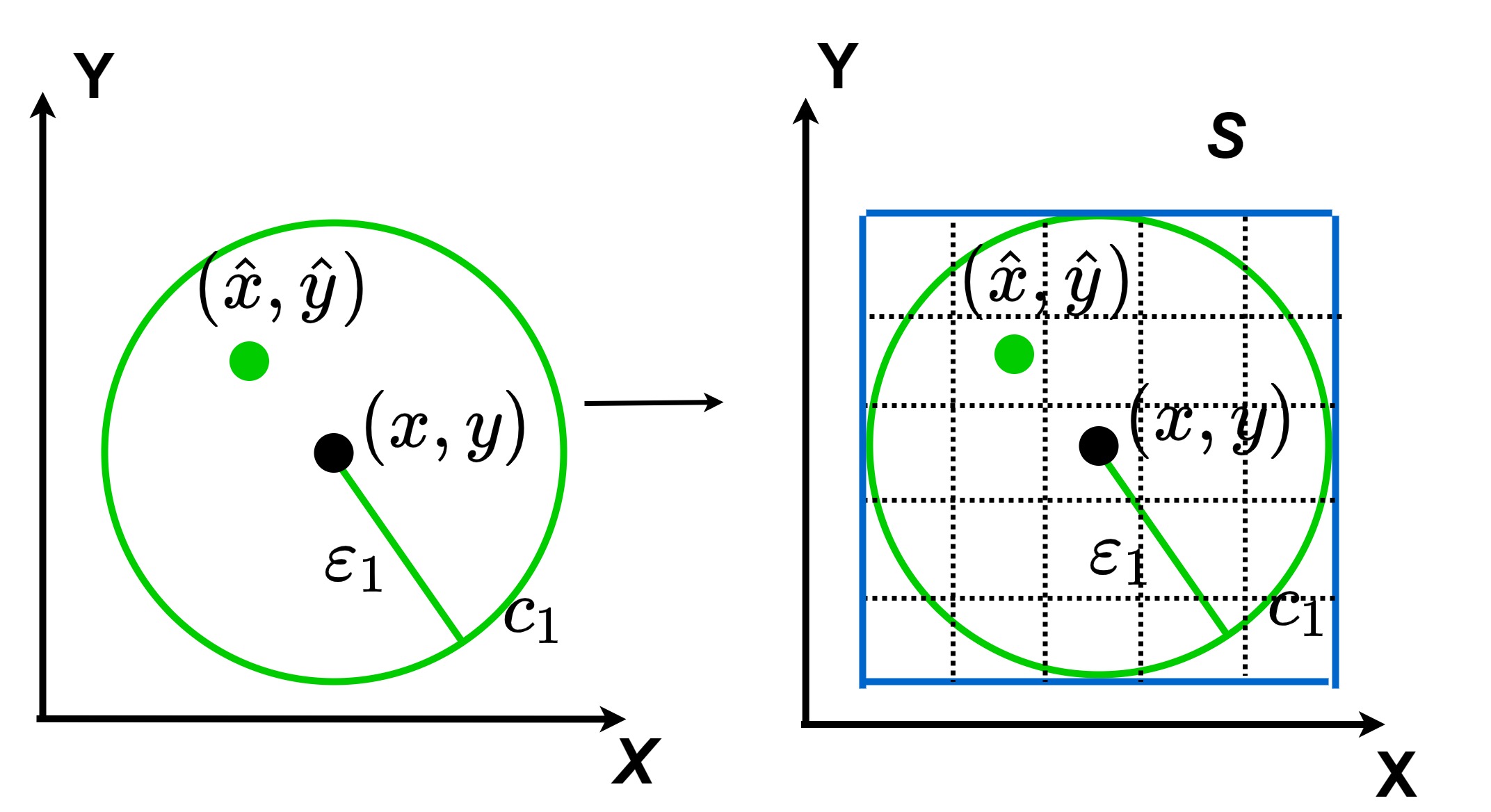}
  \caption{An error space example for the reconstructed trajectory point $(\hat{x}, \hat{y})$}
    \label{error_space}
\end{figure}

With the error-bounded codebook, the reconstructed value $(\hat{x}, \hat{y})$ is guaranteed to be within the circle $c_{1}$, as shown in Figure \ref{error_space}.  While a small $\varepsilon_1$ is desirable for the accuracy of approximate query results, an excessively small $\varepsilon_1$ would degrade the effectiveness of quantization, both in terms of the efficiency of learning and the size of the codebook.  Here, we present the coordinate quadtree coding (CQC), which encodes the spatial deviation between $(x, y)$  and $(\hat{x}, \hat{y})$, to reduce the information loss of the summary.  

\begin{figure}[tb]
 \subfloat[]
 {\includegraphics[height=3.3cm,width=3.3cm]{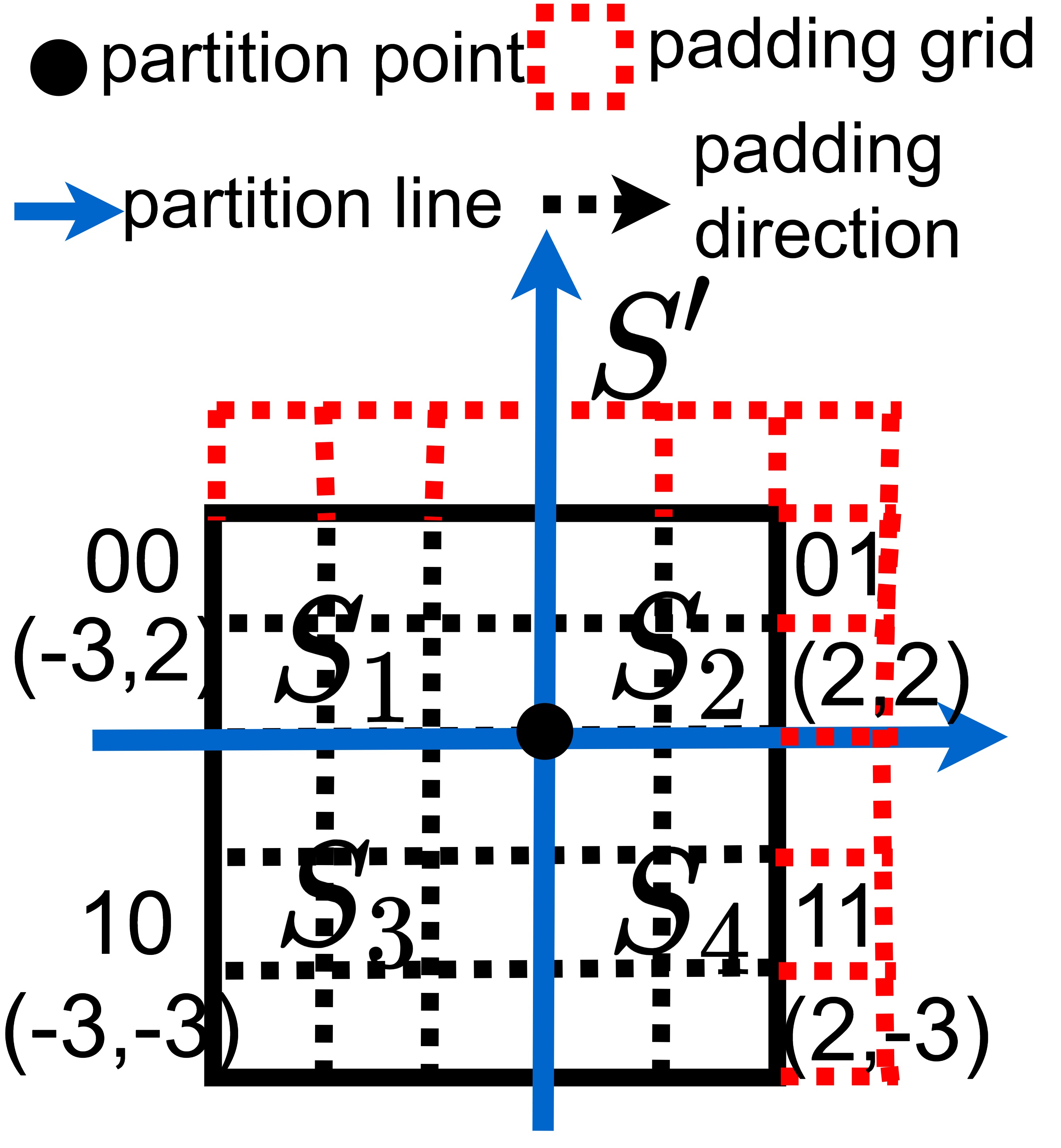}\label{padding_example_a}}
 \hspace{2mm}
 \subfloat[]
  {\includegraphics[height=3.3cm,width=3.3cm]{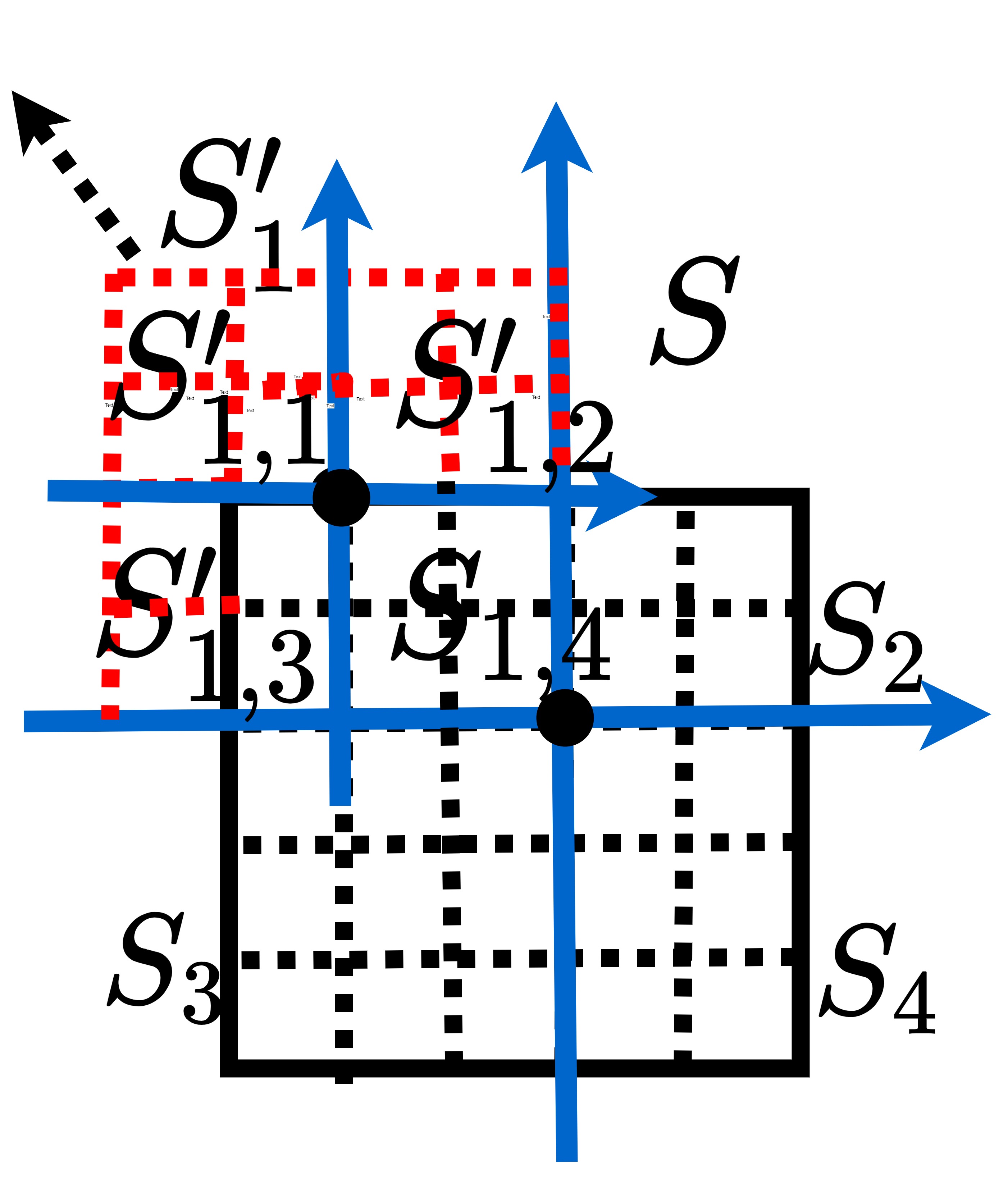}\label{padding_example_b}}
 \centering
 
 \subfloat[]
{\includegraphics[height=3.3cm,width=3.3cm]{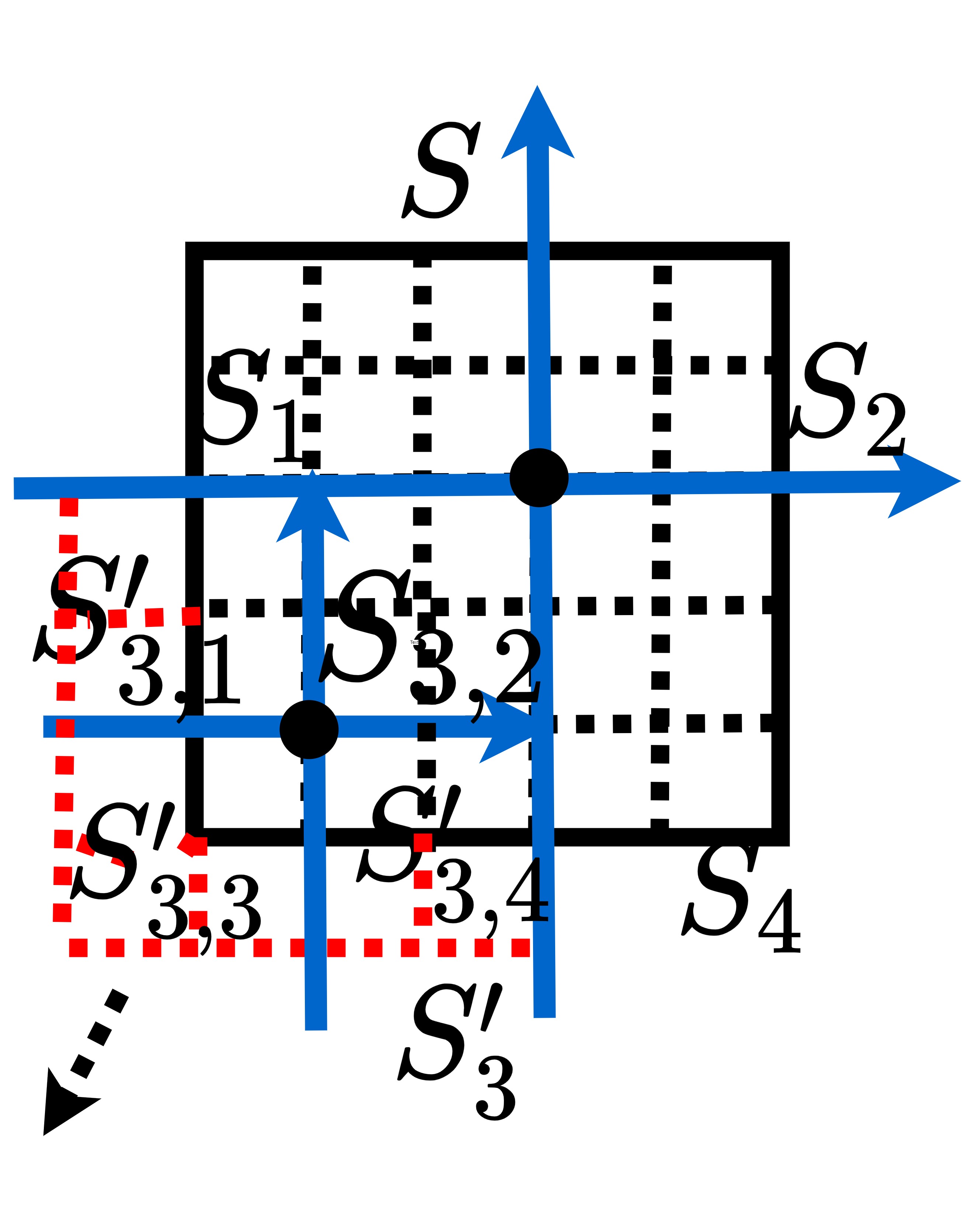}\label{padding_example_c}}
 \hspace{2mm}
 \subfloat[]
 {\includegraphics[height=3.3cm,width=3.3cm]{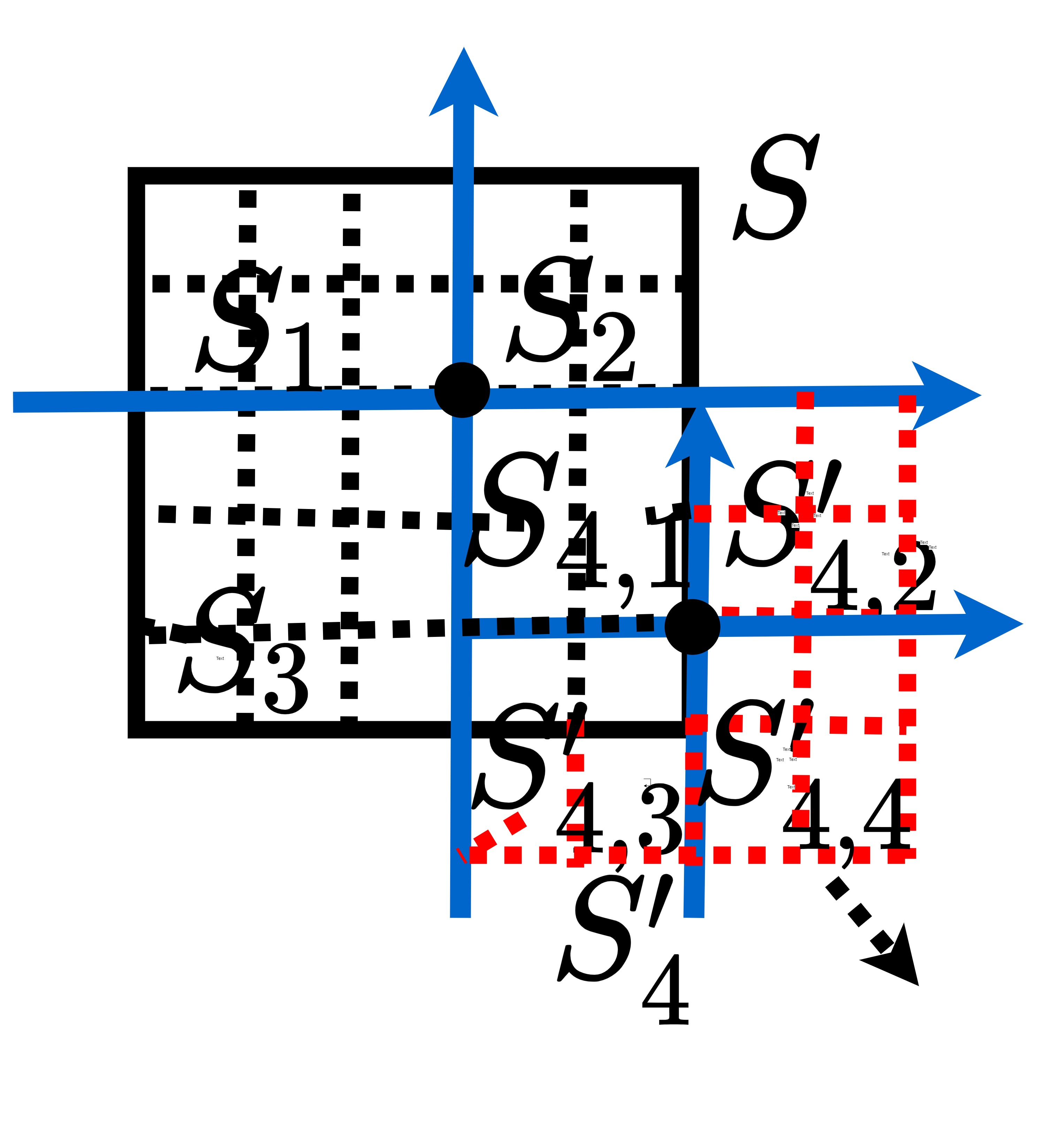}\label{padding_example_d}}
 \centering
 \caption{Padding Example}\label{padding_example_list}
 \end{figure}

CQC consists of short binary codes that can be easily restored to the relative position between $(x, y)$  and $(\hat{x}, \hat{y})$ to obtain an accurate trajectory reconstruction $(\hat{x}', \hat{y}')$. The construction of the coordinate quadtree and getting the CQC are independent of the dataset size when $\varepsilon_1$ and $g_s$ are fixed.    
\begin{definition}(Coordinate Quadtree)\label{coordinate_quadtree_definition}
A coordinate quad-tree is a tree structure in which each internal node has four children nodes, the value of a node is the coordinate of the subspace that the node represents, and the value over the edge is the quadrant that its parent node is located in. 
\end{definition}
 \begin{definition}(Coordinate Quadtree Coding)\label{define_HQC}
 Given a coordinate quadtree, 
the coordinate quadtree coding (CQC) of a node $n_q$ is the values of the edges from the root node to node $n_q$ as well as the quadrant that $n_q$ is located in.
\end{definition}

\begin{algorithm}[tb]  
	\renewcommand{\algorithmicrequire}{\textbf{Input:}}
	\renewcommand{\algorithmicensure}{\textbf{Output:}}
	\caption{Coordinate\_QuadTree($c_{1}$, $g_s$)}
	\label{alg:hqc}
	\begin{algorithmic}[1]
		\ENSURE $Cq$
		
		\STATE Get the minimum rectangle $S$ covering  $c_{1}$
		
		\STATE $S$ is split into grids of equal size, $S_{g_s}$.

 		       \STATE $Cq$ = $\{\}$ \# Coordinate quadtree 
 		        \STATE $build\_tree(S_{g_s}, Cq)$.
    
	\end{algorithmic}  
\end{algorithm}

\begin{algorithm}[tb]  
\renewcommand{\thealgorithm}{}
\floatname{algorithm}{Function}
\setcounter{algorithm}{0}
  	\renewcommand{\algorithmicrequire}{\textbf{Input:}}
	\renewcommand{\algorithmicensure}{\textbf{Output:}}
	\caption{$build\_tree(\{S_{g_s,i}\}$, Cq)}
	\label{alg:build_tree}
	\begin{algorithmic}[1]
		\ENSURE $Cq$. 
		  \FORALL{each sub-region $S_{g_s}$ in  \{$S_{g_s,i}$\}}
		           	\STATE $s_x$, $s_y$ $\leftarrow$ $|S_{g_s}|$ \# $s_x$ and $s_y$ are the number of grid cells of $S_{g_s}$ along the $x$- and $y$-axes respectively
		  
		            \IF{($s_x,s_y=1$$  \&  no\_padding(S_{g_s})$) or $s_x,s_y$ = 0}
		                   \STATE continue
		             \ENDIF
             		      \STATE \{$\dot{S}_{g_s,i}$\}$\leftarrow$partition\_padding($S_{g_s}$)   
    		         \STATE   $Cq.append( \{\dot{S}_{g_s,i}\})$.
		            
    		        \STATE      $build\_tree(\{\dot{S}_{g_s,i}\}, Cq)$
		  \ENDFOR
	\end{algorithmic}  
\end{algorithm}

 \begin{algorithm}[tb]  
 \renewcommand{\thealgorithm}{}
 \floatname{algorithm}{Function}
   	\renewcommand{\algorithmicrequire}{\textbf{Input:}}
 	\renewcommand{\algorithmicensure}{\textbf{Output:}}
 	\caption{$partition\_padding$($S_{g_s}$)}
 	\label{alg:build_tree}
 	\begin{algorithmic}[1]
 		\ENSURE $\{S_{g_s,i}\}$. 
		\IF{$\lfloor\frac{s_x}{2}\rfloor \ne \frac{s_x}{2}$ or $\lfloor\frac{s_y}{2}\rfloor  \ne \frac{s_y}{2}$}
             		      \STATE $S_{g_s}'\leftarrow$padding($S_{g_s}$)
             		       \STATE Partitioning $S_{g_s}'$into \{$S_{g_s,i}$\}
             		           
    	 \ELSE
    	      \STATE Partition $S_{g_s}$ into \{$S_{g_s,i}$\} \# four equal partitions
    	 \ENDIF
		\end{algorithmic}  
 \end{algorithm}

\subsection{Coordinate Quadtree Coding}

\begin{figure}[tb]
\subfloat[Coordinate Quadtree Partitioning Process]
{\includegraphics[height=3.9cm,width=7.5cm]{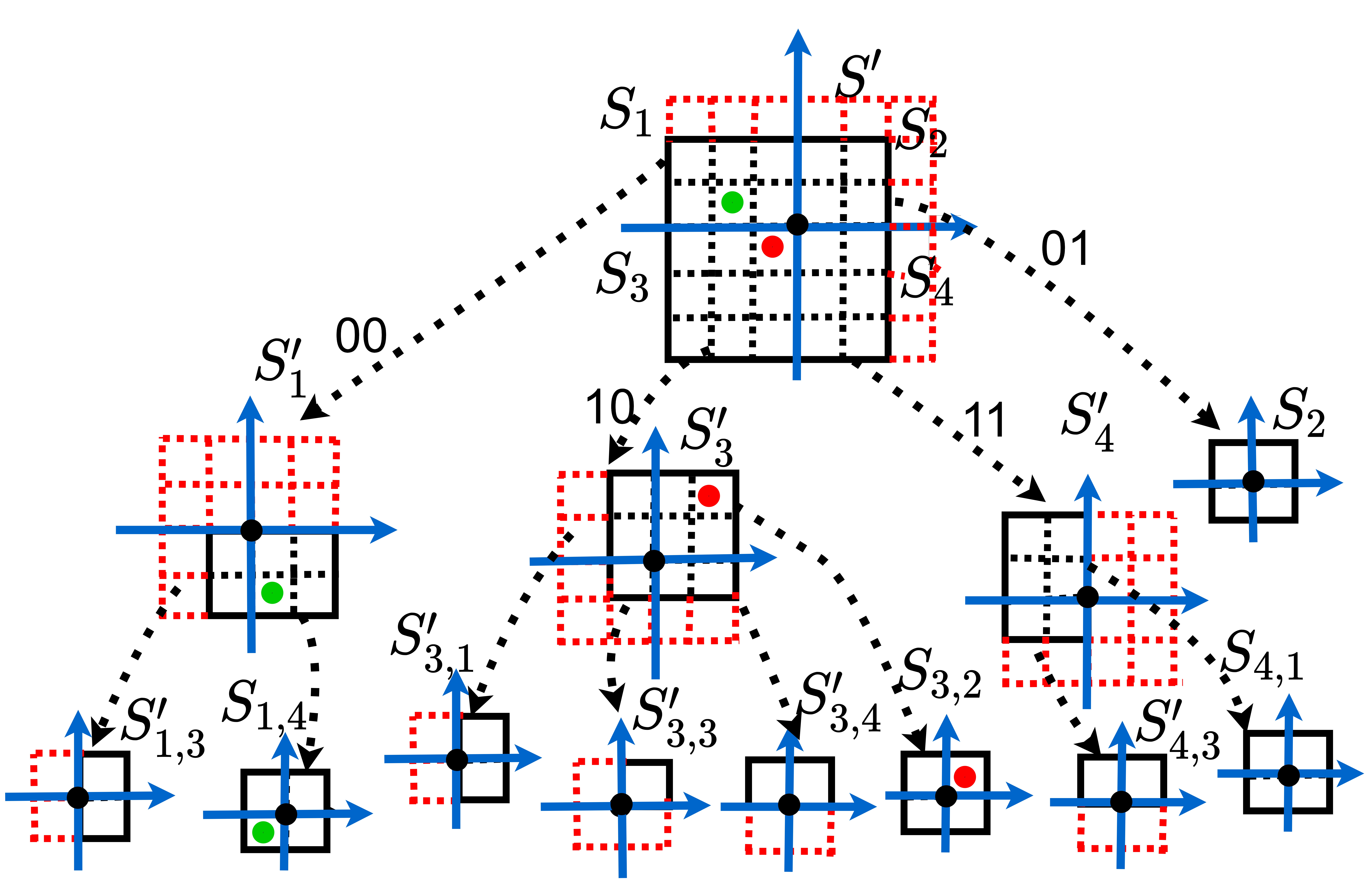}\label{whole_padding_process}}

\subfloat[Coordinate Quadtree]
{\includegraphics[height=3.9cm,width=7.5cm]{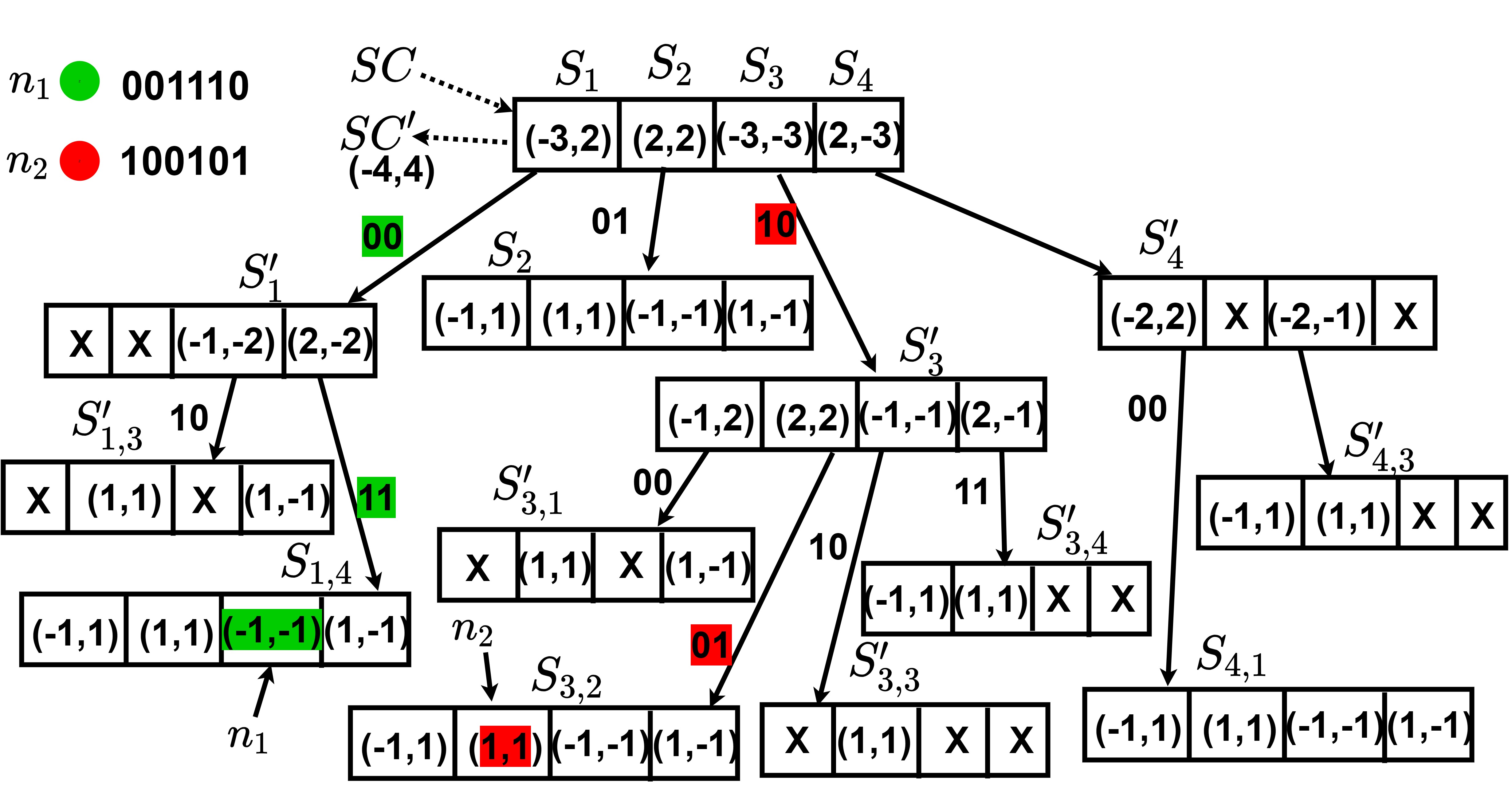}\label{coordinate_quadtree}}
\caption{Coordinate Quadtree Coding Example}
\label{allHQI}
\end{figure}

The process of building the coordinate quadtree, which is used as the basis for CQC, is summarized in Algorithm \ref{alg:hqc}.
The first step is to get the error space $c_{1}$ and find the minimum rectangle $S$ covering $c_{1}$ in Line 1.
The second step is 
to divide $S$ into grid cells of equal size, $S_{g_s}$, 
in Line 2, where $g_s$ is the size of a cell.
The third stage is to build the coordinate quadtree via Function $build\_tree$. 
Its stopping condition is either when the subspace is empty, or the input subspace is size one and without any padding grid cells, as shown in Line 3. For any $S_{g_s}$, its partitions are generated by $partition\_padding$ in Line 4.  An example is given in Figure \ref{padding_example_list}. For simplicity, we omit $g_s$ and use $S_{i}'$ and $S_{i}$  to demonstrate the example. 
For function $partition\_padding$, if $S$ does not satisfy Line 1,  we pad $S$ as $S'$ so that $S'$ can be partitioned into four equal partitions.  To avoid conflicts of partitions at different rounds, we design specific padding rules for different quadrants.  We have a 5 $\times$ 5 grid $S$ in Figure \ref{padding_example_a}, $S$ is expanded to a 6 $\times$ 6 
grid $S'$ to obtain four equal size subspaces. 
According to Definition \ref{coordinate_quadtree_definition}, the values (-3,2), (2,2), (-3,-3) and (2,-3) represent the size and quadrant for every subspace. The quadrants are encoded as $00$, $01$, $10$ and $11$ separately.
As shown in Figure \ref{padding_example_b}, the subspace $S_1$ at quadrant $00$ is expanded towards the upper left as $S_1'$ which can be further partitioned. Similarly, the subspaces at quadrant $10$ and $11$, i.e., $S_3$ and $S_4$, are padded towards the bottom left and bottom right respectively, as shown in Figure \ref{padding_example_c} and Figure \ref{padding_example_d}. However, there is no need to pad $S_2$ as it can be directly partitioned into four subspaces of equal size following Line 5. 
The full process of building a coordinate quadtree for 5 $\times$ 5 grids is shown in Figure \ref{whole_padding_process}. The corresponding coordinate quadtree can be observed in Figure \ref{coordinate_quadtree}. The value in every square corresponds to the coordinate of a subspace it denotes, which is denoted as $SC$. $``X"$ denotes the empty padding grid. According to Definition \ref{define_HQC}, the CQC for the 
$n_1$ in Figure \ref{allHQI} is $001110$. 

For a CQC, the real coordinate is the sum of the values of the nodes it visits, as shown in Equation \ref{eq:hqc_cal}. 
\begin{equation}
c_{cqc} =  \sum_{j=1}^{m} \frac{1}{2}{ SC_j' } 
\label{eq:hqc_cal}
\end{equation}
\begin{equation}
SC'=
\begin{cases}
SC&\left|x\right|=\left|y\right|=1\\
2 \left \lceil \frac{\max ( \left| x \right|, \left| y \right|)}{2} \right \rceil \cdot (\text{sgn}(x), \text{sgn}(y)) & \text{otherwise}
\end{cases} \label{eq:pad_LQC}
\end{equation}
where $SC'$ denotes the $SC$ of the padded subspace. 
$SC'$ is obtained by  Equation \ref{eq:pad_LQC}. For example, in Figure \ref{coordinate_quadtree}, a $SC$ is (-3,2), its $SC'$ is (-4,4) by Equation \ref{eq:pad_LQC}. According to Equation \ref{eq:hqc_cal}, the real coordinate for $n_1$ is ($\frac{-3}{2}, \frac{1}{2}$).

If $S_{g_s}$ cannot be partitioned into four subspaces with the same number of grid cells, the traditional approach for quadtrees \cite{samet1984quadtree} would extend $S_{g_s}$ and maintain the empty grids.
However, 
the padding process guarantees to produce four equally sized partitions. We utilize the coordinate of the subspace, to keep track of the real size of every subspace and the relative displacement of grid cells among the subspaces at different levels,
and make sure one can reversely restore the real position of any grid cell in $S_{g_s}$ by Equation \ref{eq:hqc_cal}--\ref{eq:pad_LQC}.

\subsection{Trajectory Reconstruction with CQC} \label{trajectory_reconstruction_with_CQC}

When $\varepsilon_1$ and $g_s$ are given, ($x$, $y$) is fixed at the center cell of $S_{g_s}$, i.e, $n_2$ in Figure \ref{allHQI}. Its CQC is represented as $cqc_{1}$. The CQC of $(\hat{x}, \hat{y})$ is denoted as $cqc_{2}$. With  $cqc_{1}$ and $cqc_{2}$, the reconstructed trajectory point for $(x,y)$ is obtained as:
\begin{equation}
(\hat{x}', \hat{y}') = (\hat{x}, \hat{y}) + g_{s}\cdot (c_{cqc_1} - c_{cqc_2}) 
\label{eq:coordinate_recover}
\end{equation}
where $c_{cqc_1}$ and $c_{cqc_2}$ are CQC of $(x,y)$ and $(\hat{x},\hat{y})$, obtained by Equation \ref{eq:hqc_cal}.

As shown in Figure \ref{coordinate_quadtree}, $(x,y)$ is denoted as $n_2$,  $(\hat{x},\hat{y})$ is represented as $n_1$. According to Equation \ref{eq:hqc_cal}, the real coordinates of $(x,y)$ and $(\hat{x},\hat{y})$ at the $S_{g_s}$ are $c_{cqc_1}$ = $(-\frac{1}{2}, -\frac{1}{2})$ and $c_{cqc_2}$ = $(-\frac{3}{2}, \frac{1}{2})$, respectively. Then, when $(\hat{x}, \hat{y})$ is known, the reconstructed point, $(\hat{x}', \hat{y}')$, can be obtained using Equation \ref{eq:coordinate_recover}.
Given 
$\varepsilon_1$ and $g_s$, a unified and fixed coordinate quadtree is obtained. 
The coordinate quadtree is stored as a template to recover the position with the given CQC.
For any original trajectory points $(x,y)$, they have the same CQC, i.e., ${cqc_1}$, as they are always located at the center of the same grid of $S_{g_s}$. 
Hence, only the coding of $(\hat{x},\hat{y})$, i.e.,  $cqc_{2}$, is stored for every trajectory point sample. $(\hat{x}, \hat{y})$ is recovered from the summary produced by our approach.
It is easy to prove that deviation has been reduced within $\frac{\sqrt{2}}{2}g_{s}$ as shown in Lemma \ref{HQC_error_analysis}. 
 \begin{customlemma}{3}\label{HQC_error_analysis}
    Given the  grid size $g_{s}$, the trajectory point $(x,y)$, the error of the accurate reconstructed trajectory $(\hat{x}', \hat{y}')$ does not exceed $\frac{\sqrt{2}}{2}g_{s}$,  i.e., $ 0 \leq\left\|(x,y) - (\hat{x}', \hat{y}')\right\|_2 \leq \frac{\sqrt{2}}{2}g_{s}$.  
    \end{customlemma} 
 \begin{proof}
  As shown in Figure \ref{error_space},  the continuous subspace is quantized into the grid cell of size $g_s$. Any points falling into the same grid cell are quantized into the same value, e.g., the center of that grid cell. Hence, the maximum error between the quantized value and the real value is $ \frac{\sqrt{2}}{2}g_{s}$, i.e., half of the length of the grid diagonal. According to Equation \ref{eq:coordinate_recover}, the deviation over the quantized values has been kept by CQC. With CQC, the unmeasured error is just the deviation introduced by the quantized process of ($\hat{x}, \hat{y}$), i.e., the maximum is $\frac{\sqrt{2}}{2}$ $g_{s}$. Hence, we get $ 0 \leq\left\|(x,y) - (\hat{x}', \hat{y}')\right\|_2 \leq \frac{\sqrt{2}}{2}g_{s}$. 
 \end{proof}  

\section{Online Querying over Quantized Trajectories}
\label{TSDI}
The last part of PPQ-trajectory is the organization of quantized spatio-temporal data for online querying to obtain the candidate set without the reconstruction of all trajectories. A simple spatio-temporal query example is to search vehicles that travel through location ($x$, $y$) at time $t$, and estimating their next $l$ positions. We focus on two commonly used spatio-temporal queries, as presented later in the paper. 

Since the parameters in the system ($\{P_j[t]\}$, $C$, $\{b^{t}_i\}$, CQC) are enough to reproduce any trajectory, the naive solution is to reconstruct the trajectories from time 1 to $t$ and return the trajectories that match the query conditions. In Section \ref{time_series_partition_based_index}, we propose a temporal partition-based organization to enable direct access to the relevant trajectory IDs at time $t$, and get their next $l$ positions in an online manner. In Section \ref{query_over_gmi}, we illustrate the spatio-temporal query process over quantized trajectories, and a local search strategy motivated by CQC is introduced to achieve the accurate query.

\subsection{Temporal Partition-based Index} 
\label{time_series_partition_based_index}
\begin{figure}[tb]
 \subfloat[A PI example at $t$]
 {\includegraphics[height=2.3cm,width=8cm]{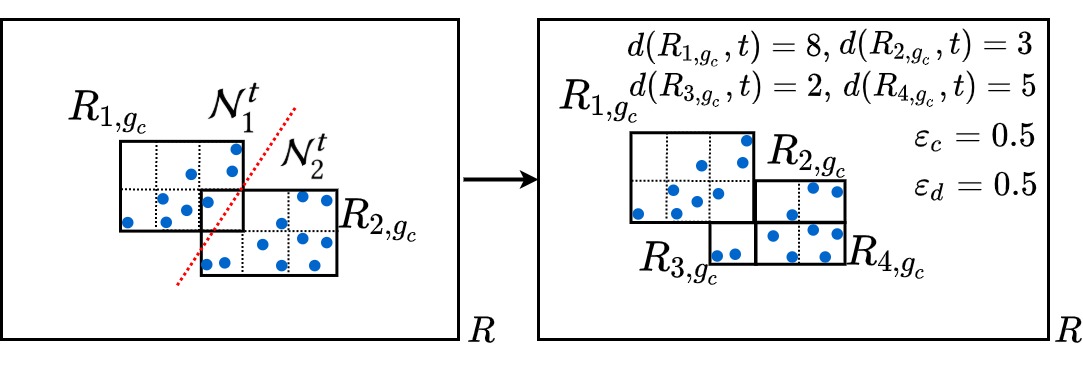}\label{an_PI_example}}

 \centering
 \subfloat[``Re-build'' case at $t+1$]
 {\includegraphics[height=2.3cm,width=8cm]{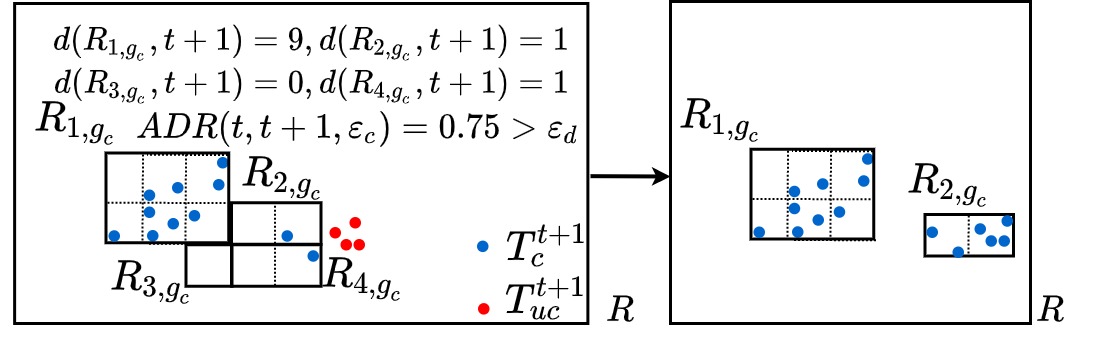}\label{an_rebuilt_example}}
 
 \centering
 \subfloat[``Insertion'' case at $t+1$]
 {\includegraphics[height=2.3cm,width=8cm]{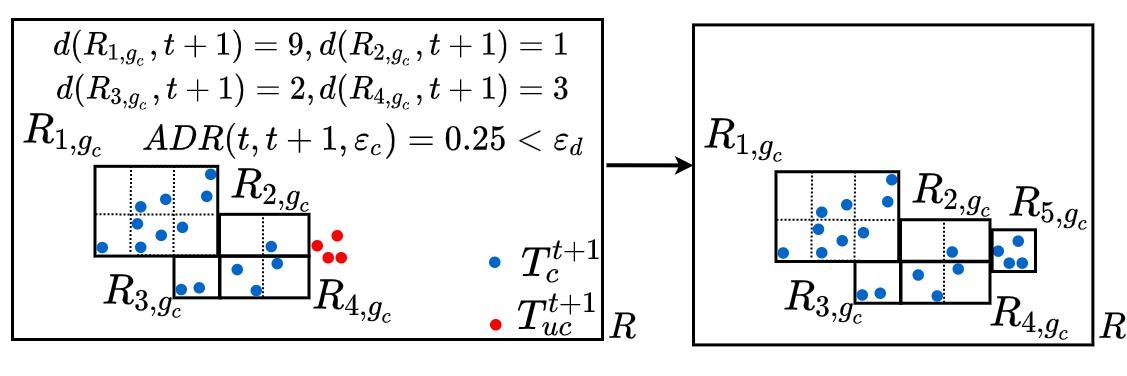}\label{an_insertion_example}}
 \caption{An illustrative example of TPI}\label{A_TPI_Illustrative_Example}
 \end{figure}

Given the non-uniform nature of trajectory data, a temporal partition-based index (TPI) over the quantized trajectories is constructed to prune the search space. TPI can actually be applied for any of ${T_i}^t$, ${{\hat{T_i}}^t}'$and ${\widehat{T_i}}^t$, for simplicity, 
we use ${T_i}^t$ to illustrate, which is interchangeable with ${{\hat{T_i}}^t}'$ and ${\widehat{T_i}}^t$. The process of constructing the partition-based index (PI), at time $t$,  is described in Algorithm \ref{alg:DI}. 
An example for PI at $t$ is given in Figure \ref{an_PI_example}. 
In Line 1, the trajectory points at time $t$, i.e., $T^t$, are partitioned into $q_s$ subsets following the same principles as Equation \ref{eq:ppq_spatial_partition} while replacing $\varepsilon_p$ with $\varepsilon_s$, where $\varepsilon_s$ is the partition threshold for region $R$. Specifically, the setting of $\varepsilon_s$ depends on the size of the region $R$ we operate on. For every ${\mathcal{N}_j}^t$,we find the minimum rectangle $R_n$ to cover the trajectory points of ${\mathcal{N}_j}^t$, as shown in Line 5. For example, points in Figure \ref{an_PI_example} are split into two partitions, i.e.,${\mathcal{N}_1}^t$ and ${\mathcal{N}_2}^t$. The minimum rectangle $R_1$ is found to cover trajectory points of ${\mathcal{N}_1}^t$. Note that overlap between $R_n$s might exist.  In Lines 7--8, to avoid duplicate indexes for some points, the overlapping region is removed, then the left polygon is separated into non-overlapping rectangles by the approach in \cite{gourley1983polygon}, which is denoted by the function $remove\_overlap$. As shown in Figure \ref{an_PI_example},  $R_2$ overlaps with $R_1$, the overlapping region is removed, and the left polygon is separated into $R_2$, $R_3$ and $R_4$.
As shown in Line 11, we build a grid index \cite{wang2017answering,wang2018torch} for the set $\{R_{1}, ..., R_{n}\}$ by Function $grid\_index$. Specifically, for every subregion $R_{j}$, it is partitioned into cells of $g_c$ \cite{li2018deep}. Every trajectory point $T^t_i$ is then mapped to the corresponding grid cell, and its trajectory ID is stored into the grid cell. To reduce the storage cost, we compress trajectory IDs mapped to the grid cell by delta encoding and Huffman codes, following the approach in the other works \cite{jegou2010product,lemire2015decoding,wang2018torch}. The PI at time $t$, i.e., $pi_t$, is returned as shown in Line 12.

\begin{algorithm}[tb]
	\renewcommand{\algorithmicrequire}{\textbf{Input:}}
	\renewcommand{\algorithmicensure}{\textbf{Output:}}
	\caption{PI}
	\label{alg:DI}
	\begin{algorithmic}[1]
		\REQUIRE $T^t$, partition threshold $\varepsilon_s$, grid size $g_c$
		\ENSURE  $pi_t$
		 	\STATE Get $q_s$ partitions $\{\mathcal{N}^t_1,...,\mathcal{N}^t_{q_s}\}$ \#following Equation \ref{eq:ppq_spatial_partition} while replace $\varepsilon_p$ with $\varepsilon_s$
		 	    \STATE $region\_list$ = $\{\}$, $n$ = 0 
		       \FOR{each $\mathcal{N}^t_j$ in $\{ \mathcal{N}^t_1,...,\mathcal{N}^t_{q_s}\}$}
		    \STATE $n$ = $n$ + 1
		    \STATE Find the minimum rectangle $R_n$ covering trajectory points of $\mathcal{N}^t_j$ .
		    \IF{$R_n$ overlaps with rectangles in $region\_list$}  
		       \STATE $\{R_{n}, ..., R_{n+l-1}\}$=$remove\_overlap$($R_{n}$),
		       \STATE$region\_list$.append($\{R_{n}, ..., R_{n+l-1}\}$), $n$=$n$+$l$-1.
		    \ELSE 		
		        \STATE $region\_list$.append($R_n$)
		   \ENDIF 	   
		\ENDFOR   
	     \STATE $\{R_{1, g_c}, ... , R_{n, g_c}\}$ = $grid\_index$($\{R_{1}, ..., R_{n}\}$) 
	    \STATE \textbf{return}  $pi_t$=$\{R_{1, g_c}, ... , R_{n, g_c}\}$
	\end{algorithmic}  
\end{algorithm}

We avoid building PI from scratch for every timestamp to efficiently maintain dynamic trajectories. For example, at time $t_s$,  $T^{t_s}$ are indexed by $pi_{t_s}$. At time $t_e$, part of $pi_{t_s}$  might be reused for $T^{t_e}$, as the distributions among consecutive timestamps might not change sharply.

\begin{definition}
(Trajectory Region Density (TRD))\label{TRD_definition}
Given $T^t$
and its PI $pi_t$=$\{R_{1, g_c},...,R_{n, g_c} \}$, 
for $R_{i, g_c}$, its TRD is $d(R_{i, g_c},t)=\frac{N_{R_{i, t}}}{|R_{i, g_c}|}$,
where $|R_{i, g_c}|$ denotes the size of rectangle $R_{i}$, 
$N_{R_{i, t}}$ is the number of trajectories indexed by $R_{i, g_c}$ at time $t$.
\end{definition}

The definition of TRD quantifies the occupancy rate of subregions, providing the basis of building the temporal index flexibly. According to Definition \ref{TRD_definition}, we compute the average dropping rate (ADR) of TRD by Equation \ref{eq:dropping_rate_trd}, to measure reusing the previous index $pi_{t_s}$ or constructing a new index.
For example, at time $t_s$, its index is  $pi_{t_s}$ = $\{R_{1, g_c},...,R_{n, g_c} \}$. The TRD of subregion $R_{i, g_c}$ at time $t_s$, i.e., $d(R_{i, g_c},t_s)$, is obtained. For trajectory points at time $t_e$, the new TRD $d(R_{i, g_c},t_e)$ is computed. For $R_{i,g_c}$, the dropping rate of its TRD from $t_s$ to $t_e$ can be obtained with Equation \ref{eq:pad_LQCA1}. For $R_{i, g_c}$, if the dropping rate of its TRD, i.e., $h_1(R_{i,g_c},t_e, t_s)$, exceeds the threshold $\varepsilon_c$, then it counts for ADR since the occupancy rate of $R_{i,g_c}$ drops too much, which is achieved by Equation \ref{eq:pad_LQCA}. 
\begin{equation}
ADR(t_s,t_e,\varepsilon_c)=\mathrm{ \sum_{i=1}^{n} \frac{h(h_1(R_{i,g_c},t_e, t_s),\varepsilon_c)}{n} }\label{eq:dropping_rate_trd} \\
\end{equation}
\begin{equation}
h_1(R_{i,g_c},t_e, t_s)= \frac{d(R_{i, g_c},t_e) - d(R_{i, g_c},t_s)}{d(R_{i, g_c},t_s)}\label{eq:pad_LQCA1}
\end{equation}
\begin{equation}
h(x,\varepsilon_c)=
\begin{cases}
1&x \textless 0 \quad and  \quad |x| \textgreater \varepsilon_c  \\
0 & \text{others}
\end{cases} \label{eq:pad_LQCA}
\end{equation}

\begin{algorithm}[tb]
	\renewcommand{\algorithmicrequire}{\textbf{Input:}}
	\renewcommand{\algorithmicensure}{\textbf{Output:}}
	\caption{TPI}
	\label{alg:TSDI}
	\begin{algorithmic}[1]
		\REQUIRE Trajectory dataset $T$, density error threshold $\varepsilon_d$, partition threshold $\varepsilon_s$, TRD dropping rate threshold $\varepsilon_{c}$, grid size $g_c$
		\ENSURE Time periods $\{period_j\}$ and PIs $\{pi_{j}\}$
	
			\STATE $t_s$ = 1, $t_e$ =1, j = 1
			\STATE $pi_{j}$ = $PI(T^{t_e},\varepsilon_s, g_c)$
			\STATE $t_e$ = $t_e$ + 1. 
		\WHILE{data input at $t_e$}
            
              \STATE  $T^{t_e}$ = $T^{t_e}_{c} \cup T^{t_e}_{uc}$
		 	  \IF{ADR($t_s$,$t_e$,$\varepsilon_c$) $\textgreater$ $\varepsilon_d$}  
		          \STATE $period_j$.s = $t_s$, $period_j$.e = $t_e$ -1.
		          \STATE $j$ = $j$ + 1, $t_s$ = $t_e$.
		          \STATE $pi_{j}$ = $PI(T^{t_e},\varepsilon_s, g_c)$ \# Re-build
		    \ELSIF{$T^{t_e}_{uc}$ is non-empty} 
		        \STATE $pi_{j}$.append($PI$($T^{t_e}_{uc}$, $\varepsilon_s$, $g_c$)) \#Insertion
		    \ENDIF
            \STATE $t_e$ = $t_e$ + 1
		\ENDWHILE
	    \STATE \textbf{return}  $\{period_j\}$ and $\{pi_{j}\}$
	\end{algorithmic}  
\end{algorithm}
Algorithm \ref{alg:TSDI} presents the TPI. For $T^t$, the initial index $pi_t$ is obtained by PI (lines 1--3). As mentioned above, Figure \ref{an_PI_example} shows a PI example at time $t$. At next timestamp $t_e$, $T^{t_e}$ is partitioned into two parts, i.e., $T^{t_e}_{c}$ and $ T^{t_e}_{uc}$ (line 5), where $T^{t_e}_{c}$ is the set covered by $pi_j$, $ T^{t_e}_{uc}$ is the set that are not covered by $pi_j$. For the covered trajectory set $T^{t_e}_{c}$, if its ARD exceeds $\varepsilon_d$, the current index $pi_j$ can not index $T^{t_e}$ efficiently. Then a new PI is built for $T^{t_e}$ as shown in Line 6--9, which is denoted as ``Re-build'' in the experimental study.  
Otherwise, we only construct the new $PI$ for $T^{t_e}_{uc}$, i.e., ``Insertion'' in the experiments, and according to Line 10--11, the current index, $pi_j$, is updated.  Finally, we obtain a set of time periods $\{period_j\}$ and corresponding PIs $\{pi_j\}$. 
A larger $\varepsilon_d$ would lower the frequency of ``Re-build'', i.e., a higher tolerance for decreasing of TRD reduces ``Re-build''s. A smaller $\varepsilon_d$, the operation of ``Re-build'' will be more frequent due to the strict constraints for ADR.
An example is presented in Figure \ref{A_TPI_Illustrative_Example}, with $\varepsilon_{c}$ = 0.5, $\varepsilon_{d}$ = 0.5,
$|R_{i, g_c}|$ = 1 (in Definition \ref{TRD_definition}), and $d(R_{i, g_c},t)= {N_{R_{i, t}}}$, i.e., the number of nodes in $R_{i, g_c}$ at time $t$.
As shown in Figure \ref{an_rebuilt_example}, TRD at $t+1$ has changed, i.e., $d(R_{i, g_c},t+1)$, and according to Equation \ref{eq:dropping_rate_trd}--\ref{eq:pad_LQCA}, we get $ADR(t,t+1,\varepsilon_{c})=0.75 \textgreater \varepsilon_{d}$, i.e., the average dropping rate of TDR exceeds the threshold, which means the PI at $t$ can not be further reused to index trajectories at time $t+1$, hence a PI is rebuilt. However, in Figure \ref{an_insertion_example}, $ADR(t,t+1,\varepsilon_{c})=0.25 \textless \varepsilon_{d}$, then we can hold the PI at $t$, i.e., $\{R_{1, g_c}, R_{2, g_c}, R_{3, g_c}, R_{4, g_c}\}$, 
and only build a new PI for trajectory points that are not covered by PI at $t$, i.e.,  indexing
the uncovered trajectory points $T^{t+1}_{uc}$ in $R_{5, g_c}$, which is ``Insertion''.

The merits of TPI are two-fold. First, based on the dynamic trajectory density, spatio-temporal trajectories are indexed as a set of periods, which lowers the frequency of partitioning the spatial region. Second, with TPI, the accuracy of the partitions within a certain time period is guaranteed by the measurement of ARD.

For disk-resident data, the trajectory points within a time period can be written into several pages and the corresponding part of the summary, i.e.,  ($\{P_j[t]\}$, $C$, $\{b^{t}_i\}$, CQC), is assigned to the corresponding page.  A lightweight index for the
assigned page number is used to record the assigned pages for the trajectory points of $period_j$, and the corresponding summary, i.e., ($period_j$, starting page number, relative page number).

\subsection{Spatio-temporal Query Processing} \label{query_over_gmi}
We now present how PPQ-trajectory answers spatio-temporal queries. We illustrate our approach using Spatio-temporal Range Query (STRQ) and Trajectory Path Query (TPQ). 
\begin{definition}
(Spatio-temporal Range Query (STRQ)) Given time $t$ and location $(x,y)$,  STRQ retrieves trajectories which are located at the grid cell that $(x,y)$ is in at time $t$. 
\end{definition}

For STRQ, given a query $(x,y,t_1)$, if $t_1\in period_j$, the sub-regions of $pi_j$ are obtained, e.g., $R_{i,g_c}$. $(x,y)$ is mapped to a grid of $R_{i,g_c}$, then a list of trajectory IDs at time $t_1$ is returned.

\begin{definition}
(Trajectory Path Query (TPQ))
Given time $t$, location $(x,y)$ and the path duration $l$, the trajectory IDs of the STRQ of $(x,y,t)$ are first retrieved, then their sub-trajectories at the time interval [t,t+$l$] are returned. \label{TPQ_definition}
\end{definition}

For TPQ, given time $t$, location $(x,y)$ and path duration $l$, the list of trajectory IDs is first returned from STRQ by searching $(x,y,t)$, the next $l$ positions of the retrieved trajectories are then directly reproduced by the indexed summary.\\



\begin{figure}[tb]
 \subfloat[]
 {\includegraphics[height=2.8cm,width=3cm]{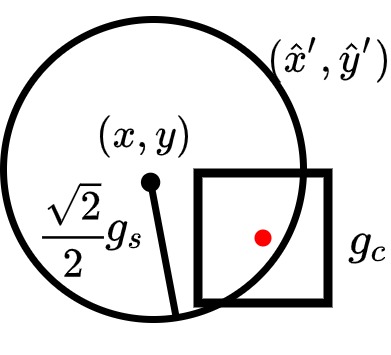}\label{grid_index_size2}}
  \hspace{7mm}
 \subfloat[]
  {\includegraphics[height=2.8cm,width=3cm]{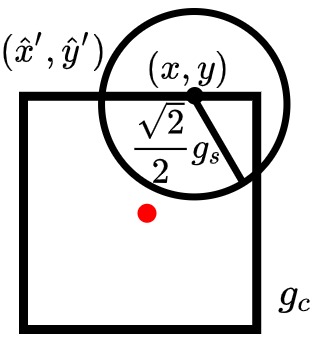}\label{grid_index_size1}}
 \centering
 \caption{Illustrating the Query Space with CQC}\label{TPQ_different_length}
 \end{figure}
\vspace{-2mm}
\noindent\textbf{Local Search using CQC.} \label{local_search} For a trajectory point $(x,y)$, its accurate reconstructed trajectory point is $(\hat{x}',\hat{y}')$. According to Lemma \ref{HQC_error_analysis},  $(\hat{x}',\hat{y}')$ might be any point within a circle  of which radius is $\frac{\sqrt{2}}{2}g_{s}$ and $(x,y)$ is the center. 
With the CQC, the deviation has been narrowed to $\frac{\sqrt{2}}{2}g_{s}$ which is a smaller distance. In order to further improve the accuracy of queries using the summary, we introduce a local search strategy. 
There are two situations for the local search. (1) When $\frac{\sqrt{2}}{2}g_{s}\textgreater g_c$,
as shown in Figure \ref{grid_index_size2}, when all the grid cells $g_c$ that are covered by the circle are scanned, the full actual result can be retrieved successfully. (2) When $\frac{\sqrt{2}}{2}g_{s}\leq g_c$, as shown in Figure \ref{grid_index_size1}, the worst case is that $(\hat{x}',\hat{y}')$ falls out of the grid cell $g_c$ as  $(x,y)$  happens to be adjacent to the border of the grid cell. Hence, in order to guarantee that the actual result is retrieved, the quantized trajectory points of which distance to $(x,y)$ is less than $\frac{\sqrt{2}}{2}g_{s}$ in the grid cells that are adjacent to the grid cell  $(x,y)$
mapped to are all scanned. Specifically, the second case is more general, i.e., $g_s$ is smaller, as the aim of introducing $g_s$ is to further reduce the information loss. For $g_c$, it serves for the index part, i.e., TPI, hence, it is usually larger. 

With the local search, the returned candidate list contains all the trajectory IDs for STRQ, which makes the recall 1. However, the candidate list may include the trajectory IDs of which true position is at the boundary of adjacent cells of $g_c$ while mapped to $g_c$.
The $(\hat{x}',\hat{y}')$  whose distance to $(x,y)$  is larger than $\frac{\sqrt{2}}{2}g_{s}$ has been filtered out according to LEMMA \ref{HQC_error_analysis}. The precision for STRQ can be improved to be 1 by accessing the original trajectory of the candidate list of which size is relatively small due to the accuracy of $(\hat{x}',\hat{y}')$.

\section{Experimental Evaluation}
\label{experiments}
We conduct a range of experiments to evaluate the effectiveness of PPQ-trajectory and compare it with a variety of alternative approaches. Our methods and all the alternative approaches are implemented in Matlab R2019a. All experiments are executed on a Ubuntu 19.04 with an Intel i5-8500 3.00 GHZ GPU and 31GB RAM. 
\subsection{Experimental Setting}\label{experiment_setting}
\noindent\textbf{Datasets}. The experiments are performed on the publicly available trajectory datasets, Porto \cite{portodataset} and GeoLife \cite{zheng2009mining}. We select the trajectories with the length being at least 30.  The selected  Porto dataset contains 1.2 million trajectories, with 74.3M trajectory points and the longest trajectory consisting of 3881 location points. The GeoLife dataset retains 17,932 trajectories with the maximum length of 92,645, containing 24.8M trajectory points.

\noindent \textbf{Compared Methods.} We implemented the extended versions of alternative methods from the literature:
Product Quantization \cite{jegou2010product}, Residual Quantization \cite{chen2010approximate}, REST \cite{zhao2018rest} and TrajStore\cite{cudre2010trajstore}. We compare with one of the variations of REST, i.e., trajectory redundancy reduction, which was shown to perform best in their work \cite{zhao2018rest}. We also test simple baselines as well as variants of PPQ-trajectory to quantify the improvements of each step.

Product Quantization and Residual Quantization are popularly used for approximate nearest neighbor (ANN) queries. However, they normally do not offer an effective index structure to support pruning for efficient querying. For fairness, we extended these methods with our indexing approach used in PPQ-trajectory.

For REST to work properly, the dataset needs to contain a highly repeating set of patterns between the trajectory set for building the reference set and the matched trajectory. This is not the case for most of the real-world data, including the
datasets we use. To be able to compare with REST, we constructed another dataset. We first randomly selected 20,000 trajectories from Porto dataset, then for every trajectory we created four more similar trajectories by down-sampling and adding noise following the procedure in \cite{li2018deep}. This process gives a dataset with 100,000 trajectories that is suitable for REST. Specifically, 2,000 trajectories are randomly selected for compression, we name this dataset as sub-Porto, while other trajectories are used to build a reference set for REST.

TrajStore builds an index with the  spatial regions and recursively updates the spatial index  by merging, splitting or appending. To align with the experimental setting, we implemented TrajStore to be able to get streaming trajectory points as input, and dynamically build the spatial index with time increasing.

We also implemented different variants of PPQ-trajectory to understand the effect of its building blocks. PPQ-S uses the spatial proximity for partitioning, PPQ-A uses the autocorrelations based similarities.  PPQ-S-basic and PPQ-A-basic use the quantizers but not CQCs. As other baselines, we include comparisons with E-PQ, and also with a basic version of PPQ-trajectory, by skipping the prediction part, and name it as Q-trajectory. 

\noindent \textbf{Parameter Settings.} The default quantization deviation threshold is 
$\varepsilon_{1}$  = 0.001, which is $\varepsilon_{1}^M \approx $ 111 meters under the geographic coordinate system
\cite{chang2008introduction}. In the following experiments, we directly use $\varepsilon_{1}^M$ to describe the comparative study. For the partition threshold $\varepsilon_p$, its setting varies on the autocorrelation and spatialproximity-based partitions. 
For the spatial proximity-based solution, $\varepsilon_p$ defaults to 0.1 for Porto and 5 for Geolife. In the case of autocorrelation similarity, $\varepsilon_p$ defaults to be 0.01 for both datasets.
For TPI and PI, the grid cell size $g_c$ is set to 100m.
$g_s$ defaults to 50m, which denotes the size of the grid cell for CQC. 
The threshold of the dropping rate of TRD, $\varepsilon_{c}$ is defaulted to be 0.5.
The default setting of the threshold of ADR, i.e., $\varepsilon_{d}$, is 0.5. $\varepsilon_{s}$ defaults to 0.1, which represents the partition threshold for constructing index.
 
 \vspace{-3mm}
 \subsection{Query Performance}

\subsubsection{Spatio-temporal Range Query}\label{spaio-temporal-range-query}
The quality of the approximate results for STRQ is evaluated in terms of precision and recall. The precision is the ratio of the correctly retrieved trajectory IDs to the returned candidate list, and the recall is the ratio of the correctly retrieved trajectory IDs to all the trajectory IDs that match the query. We also measure the MAEs of the summaries over the datasets, i.e., the mean absolute errors between the reconstructed trajectory points and the original trajectory points.

For STRQ, we learn $C$ independently for every timestamp guaranteeing the same number of codewords is given to trajectory points at the same time across all methods.  We randomly select 10,000 queries.  
The comparative results are summarized in Table \ref{compare_with_baselines1}.  
For MAE, the PPQ-trajectory significantly performs better than other methods. 
As the time increases, PPQ-trajectory is able to gradually quantize a narrower range, whereas, residual and product quantization do not improve over time.
The summary process of Trajstore cannot start until the spatial index has been updated with trajectory points of all the timestamps. To ensure fairness, the codewords are assigned in proportion to the number of trajectory points for every spatial cell of TrajStore.


With the same number of bits, PPQ-trajectory obtains significantly higher recall and precision values. 
For Geolife, autocorrelation-based partitioning (in PPQ-A and PPQ-A-basic) helps to achieve smaller MAEs compared to the spatial proximity based solution (PPQ-S and PPQ-S-basic),
while PPQ-S-basic outperforms PPQ-A-basic on Porto dataset. We observe that autocorrelation similarity possesses some advantages upon capturing correlations and obtaining a narrower dynamic range of prediction errors. This result provides a useful insight also for other partitioning tasks and applications of spatio-temporal data, which is discussed in Section \ref{partition_for_group_model}.
PPQ-S and PPQ-A use CQC, which bounds the error within $\frac{\sqrt{2}}{2}g_{s}$ as shown in  Lemma \ref{HQC_error_analysis} and reduces the MAEs. 

Using the local search strategy in Section \ref{query_over_gmi}, 
the precision and recall of PPQ-S and PPQ-A are all 1.
TrajStore has smaller MAEs and achieves higher recall and precision compared to the other baselines, as the spatially close trajectory points are finely indexed by the same cell. However, TrajStore makes use of the pre-built spatial index, and the summarization is not efficiently generated with time evolving. 

For the Geolife dataset, the MAEs of Q-trajectory, product quantization and residual quantization are extremely large, even around 20,000 meters that are unacceptable for the task of STRQ. Their corresponding precision and recall values are significantly lower than the alternatives. Hence, their precision and recall 
are marked with ``$\times$'' in Table \ref{compare_with_baselines1}. 
The large spatial region spanning
of Geolife leads to extremely large MAEs for Q-trajectory, residual quantization and product quantization.


\subsubsection{Trajectory Path Query}


TPQ involves querying timestamps and trajectory IDs of the STRQ results and reconstructing their next $10$--$50$ trajectory points. According to Definition \ref{TPQ_definition}, the retrieved sub-trajectories of TPQ depends on the returned trajectory IDs of the STRQ. 
Different methods might retrieve sub-trajectories for different trajectory IDs, 
as observed in their different recall and precision values presented in Table \ref{compare_with_baselines1}. For fairness, we select 10,000 same trajectory IDs for all the methods to measure the MAEs of the retrieved sub-trajectories by comparing each to the corresponding original sub-trajectory.


The comparative results are summarized in Table \ref{IO_access_with_different_TPQ_length}.  
The MAEs for the sub-trajectories increase with the increasing TPQ lengths, because more spatial deviations are accumulated when querying longer sub-trajectories.
The PPQ-trajectory and E-PQ significantly perform better than other methods, while the MAEs of E-PQ are smaller than that of PPQ-A-basic and PPQ-S-basic. 
The MAEs of Q-trajectory, residual quantization and product quantiztion increase significantly with the increasing query length, because their MAEs on the datasets have been extremely large (Table \ref{compare_with_baselines1}). We notice the MAEs of TrajStore over Porto get relatively large with $l$ increasing while its MAE on the full dataset is smaller (Table \ref{compare_with_baselines1}).
The codewords are assigned in proportion to the number of trajectory points for every spatial cell of TrajStore (Section \ref{spaio-temporal-range-query}), 
hence, a larger spatial cell with a smaller number of trajectory points scattering in will be assigned a smaller number of codewords, 
 then there will be larger deviations of summarizing the trajectory points of this spatial cell, even though the average deviations over all the spatial cells are smaller.

\subsubsection{Filtering for Exact Match Queries} 
We now present the average ratios of trajectories visited when the summary is used as an index for exact match queries. After pruning irrelevant data,  only a set of candidates is accessed. We randomly select 10,000 queries, and the average ratios of trajectories visited in their second step are presented. To ensure fairness, we learn $C$ independently for every timestamp guaranteeing the same importance is given to trajectory points at different times which will not influence the filtering ratios for different timestamped queries.  
Table \ref{compare_with_baselines} compares the MAE and ratios of trajectories visited for alternative approaches, varying the size of $C$ from five to nine bits.
TrajStore summarizes trajectory points within each cell of the spatial index,
while the spatial index is built with the trajectory points of all timestamps. Hence, for TrajStore, we cannot fairly summarize the trajectory points of every timestamp independently with the fixed size $C$. Hence, the comparison with TrajStore is not considered in this experiment.

PPQ-A performs the best under this performance measure. With the selected queries, it can directly access the trajectories mapped to the adjacent grid cells designed using $\varepsilon_{1}$ and $g_s$. The ratios of trajectories visited are the same with different sizes of $C$ due to the accurate reconstructed representation. The same applies to PPQ-S. 
For other methods, we notice their ratios gradually decrease with the size of $C$ increasing, because the accuracy that $C$ can provide increases, which helps filter more candidate results. Similar performance is observed for Geolife, especially, we observe the same average ratios of trajectories visited for PPQ-S and PPQ-A. The corresponding MAE is also presented in Table \ref{compare_with_baselines}. MAE decreases with the number of bits increasing for most of the methods. However, the MAEs of PPQ-A and PPQ-S do not strictly decline with the size of $C$ increasing, because their spatial deviation is not fully decided by the quality of $C$, but also slightly influenced by CQCs.
\subsection{Building Time Efficiency}



\begin{table}[]
\caption{Quality of summaries and STRQ evaluation} 
	\label{compare_with_baselines1}  
 \scriptsize
\begin{tabular}{ccccccc}
\hline
Dataset               & \multicolumn{3}{c}{Porto}    & \multicolumn{3}{c}{Geolife} \\ \hline
Performance Measure   & MAE(m)  & Precision & Recall & MAE(m) & Precision & Recall \\ \hline
PPQ-A                 & \textbf{18.35}   & \textbf{1.000}     & \textbf{1.000}  & \textbf{4.85}   & \textbf{1.000}     & \textbf{1.000}  \\ 
PPQ-A-basic           & 51.92   & 0.951     & 0.948  & 6.17   & 0.987     & 0.987  \\ 
PPQ-S                 & 23.30   & \textbf{1.000}     & \textbf{1.000}  & 7.89   & \textbf{1.000}     & \textbf{1.000}  \\
PPQ-S-basic           & 44.41   & 0.944     & 0.939  & 14.72  & 0.976     & 0.976  \\ 
E-PQ                  & 76.60   & 0.931     & 0.926  & 15.06  & 0.962     & 0.961  \\
Q-trajectory          & 1752.29 & 0.425     & 0.427  & 29105  & x         & x      \\ 
Residual Quantization & 868.96  & 0.675     & 0.675  & 22590  & x         & x      \\ 
Product Quantization  & 641.34  & 0.736     & 0.725  & 21228  & x         & x      \\ 
TrajStore             & 152.13 & 0.917     & 0.919  & 617.76  & 0.8535         & 0.8547      \\ \hline
\end{tabular}
\end{table}


 \begin{table}[]
\caption{MAE against different lengths of TPQ ($1.0e^{3}$m)} 
	\label{IO_access_with_different_TPQ_length}  
 \scriptsize
 \setlength{\tabcolsep}{1mm}{\begin{tabular}{ccccccccccc}
\hline
Dataset               & \multicolumn{5}{c}{Porto}                                                            & \multicolumn{5}{c}{Geolife}                                               \\ \hline
TPQ length ($l$)           & 10                      & 20                      & 30                      & 40  & 50  & \multicolumn{1}{l}{10} & 20   & 30   & 40   & 50                        \\ \hline
PPQ-A                 & \textbf{0.046}                      & \textbf{0.081}   & \textbf{0.111}   & \textbf{0.136}   & \textbf{0.158} & \textbf{0.011}
                      & \textbf{0.021}                      & \textbf{0.031}                      & \textbf{0.040}  & \textbf{0.050}                         \\ 
PPQ-A-basic           & 0.357                     & 0.657                      & 0.935                      & 1.194  & 1.437  & 0.073                      & 0.139   & 0.205   & 0.271   & 0.337                        \\ 
PPQ-S                 & 0.160                    & 0.295                    & 0.402                   & 0.491  & 0.566 & 0.019                     & 0.037   & 0.054   & 0.070   & 0.086                        \\ 
PPQ-S-basic           & 0.338                      & 0.623                      & 0.890                      & 1.140  & 1.374  & 0.135                      & 0.257   & 0.378   & 0.500   & 0.621                        \\ 
E-PQ                  & 0.068                      & 0.119                     & 0.162                      & 0.198  & 0.229  & 0.031                      & 0.059   & 0.086   & 0.113  & 0.139                        \\ 
Q-Trajectory          & {24.90} & {44.55} & {62.02} & 77.54  & 91.75  & 190.2                     & 360.5  & 530.0  & 697.5  & 861.1                     \\
\hline
\end{tabular}}
\end{table}
\input{exact_query_result.tex}
 \begin{table}[tb]
\caption{Running time against different spatial deviation (s)}\label{running_time_with_different_threshold}
\scriptsize
	\setlength{\tabcolsep}{0.8mm}{
	\begin{tabular}{ccccccccccc}
\hline
Dataset               & \multicolumn{5}{c}{Porto}             & \multicolumn{5}{c}{Geolife}           
                               \\ \hline
spatial deviation (m)    & 200    & 400   & 600   & 800   & 1000  & 200   & 400   & 600   & 800   & 1000                      \\ \hline
PPQ-A                 & 802.0 & 579.3 & 486.8 & 445.3 & 417.8 & \textbf{706.0} & \textbf{557.0} & 534.7 & \textbf{477.5} & \textbf{345.6}                     \\
PPQ-A-basic           & 1617  & 1502  & 1405  & 1399  & 1316  & 1201  & 678.9 & 533.8 & 525.7 & 518.1                     \\
PPQ-S                 & \textbf{633.4} & \textbf{428.5} & \textbf{381.2} & \textbf{367.8} & \textbf{348.1} & 799.5 & 591.1 & \textbf{518.7} & 504.0 & 445.5                     \\
PPQ-S-basic           & 1663  & 956.2 & 547.3 & 511.7 & 439.5 & 1244  & 1102  & 980.5 & 729.5 & 644.3                     \\
E-PQ                  & 6543  & 3110  & 1775  & 1503  & 1157  & 647.4 & 597.9 & 550.8 & 497.6 & 459.0                     \\
Q-Trajectory          & 16027 & 7530  & 4347  & 3100  & 2789  & 10698 & 8387  & 6168  & 4692  & 4152                      \\
Residual Quantization & 4765  & 2557  & 1727  & 1269  & 1094  & 13199 & 8489  & 6876  & 5538  & 4655                      \\
Product Quantization  & 4883  & 2870  & 2753  & 2667  & 2353  & 21113 & 8728  & 4181  & 3354  & 3073                      \\
TrajStore             & 12826  & 7448  & 5908  & 5355  & 4870  & 44588 & 33058 & 29261 & 27835 & \multicolumn{1}{l}{27063} \\ \hline
\end{tabular}}
\end{table} 
\subsubsection{Summary Efficiency}
\label{summary_efficiency}
We evaluate the running times of generating the summary for different solutions with spatial deviations as 200m, 400m, 600m, 800m and 1000m. According to Lemma \ref{HQC_error_analysis}, the spatial deviation of PPQ-A and PPQ-S are $\frac{\sqrt{2}}{2}g_{s}$. In the experiment, we set $\varepsilon_{1}^M=2g_{s}$ for PPQ-A and PPQ-S. For the other methods, the spatial deviation of their summary is simply determined by $\varepsilon_{1}^M$.

The building time is summarized in  Table \ref{running_time_with_different_threshold}, which gradually decreases as the spatial deviation increases. This is because the quantization process finishes with fewer iterations when the error is larger. 
 The running times of PPQ-trajectory are much smaller than 
those of Q-trajectory, residual quantization, product quantization, and TrajStore.
In our solution, the dynamic range of the prediction errors that needs to be quantized is decreasing with time $t$ evolving, hence its running time is smaller. PPQ-A and PPQ-S are more efficient  than PPQ-A-basic and PPQ-S-basic, respectively on both datasets, because for the same spatial deviation, the setting of $\varepsilon_{1}^M$ for PPQ-A and PPQ-S is larger than that of PPQ-A-basic and PPQ-S-basic, which needs fewer iterations to obtain the summary. 
For Porto dataset, the running time of our solution is up to 25 times faster than residual quantization, product quantization, Q-trajectory, and TrajStore, while E-PQ is faster than PPQ-A-basic when the spatial deviation is larger.
The running time of E-PQ is comparatively high on Porto, which is even larger than residual quantization. It shows the E-PQ cannot work as efficiently as our solutions on the large datasets, because E-PQ executes one prediction on the whole datasets, the prediction errors will be larger and need more iterations to satisfy the spatial deviation requirement. 
For Geolife dataset, the running time of our solution is 4-78 times faster than residual quantization, product quantization, Q-trajectory, and TrajStore, 
while E-PQ is slightly faster than PPQ-trajectory for some spatial deviations.  The running time of residual quantization, product quantization and Q-trajectory drops quickly with the spatial deviation increasing. However, their running time is extremely high  when the spatial deviation is 200m, as the time span of Geolife is relatively large, 
which needs even more iterations to summarize.

The running time of TrajStore is extremely high for all spatial deviations. Its running time includes both the time of building the spatial index and compression for every spatial cell,  because TrajStore depends on the spatial index to conduct the summarization, 
and the process of building the index is time-consuming due to the frequent merging, splitting, and appending operations.

 \begin{table}[tb]
	\caption{Number of codewords in $C$ against different spatial deviation ($\times 10^4$)}  
	\label{number_of_codewords}  
	\scriptsize
	\setlength{\tabcolsep}{0.8mm}{
\begin{tabular}{lllllllllll}
\hline
\multirow{2}{*}{Dataset} & \multicolumn{5}{c}{Porto(m)}                                                     &   \multicolumn{5}{c}{Geolife(m)}                                            \\ \cline{2-11} 
                         & 200            & 400            & 600            & 800            & 1000           & 200   & 400            & 600            & 800            & 1000           \\ \hline
PPQ-A            & \textbf{0.283} & {0.172}          & 0.113 & \textbf{0.088} & \textbf{0.069}         & \textbf{0.375}          & \textbf{0.264}          & \textbf{0.218}          & \textbf{0.181}          &\textbf{0.156}          \\ \hline
PPQ-A-basic       & 0.651 & 0.371 & 0.279          & 0.227          & 0.183       & 0.560          & 0.422          & 0.349          & 0.324          & 0.288                    \\ \hline
PPQ-S       & 0.284 & \textbf{0.148}          & \textbf{0.112}          & 0.093          & 0.082              & 0.487          & 0.311          & 0.242          & 0.208          & 0.183                   \\ \hline
PPQ-S-basic        &  0.877 & 0.415          & 0.265          & 0.198          & 0.162     & 0.927 & 0.605 & 0.475 & 0.390 & 0.344          \\ \hline
E-PQ        & 3.182 & 1.457          & 0.892          & 0.648          & 0.500             & 0.804          & 0.530          & 0.401          & 0.325          & 0.280                    \\ \hline
Q-trajectory      & 16.37 & 7.689          & 4.667          & 3.275          & 2.501        & 29.66          & 16.02          & 11.16          & 8.646          & 7.157                   \\
\hline
\end{tabular}}
\end{table}

\begin{table}[]
 \caption{Statistics of TPI on different $\varepsilon_{c}$} 
 	\label{TrajStore_index_TPI_PI_varepsilon_c}  
 \scriptsize
\begin{tabular}{ccccccccc}
\hline
\multicolumn{1}{l}{} & \multicolumn{2}{c}{Index Size(MB)} & \multicolumn{2}{c}{Time Cost} & \multicolumn{2}{c}{No.Periods} & \multicolumn{2}{c}{No.Insertions} \\ 
$\varepsilon_{c}$             & Porto           & Geolife          & Porto         & Geolife        & Porto         & Geolife        & Porto          & Geolife          \\ \hline
0.2                  & 863.1             & 250.0          & 1346           & 7003         & 1245           & 14627         & \textbf{4367}             & \textbf{71448}           \\ 
0.4                  & 860.1             & 241.6          & 544           & 3792          & 656           & 10100          & 7207            & 89492           \\ 
0.6                  & 859.4            & 237.6           & 458           & 3028          & 485            & 7117          & 7198            & 95308           \\ 
0.8                  & \textbf{859.1}             & \textbf{237.3}          & \textbf{418}           & \textbf{2935}          & \textbf{421}            & \textbf{6876}          & 6637           & 101187          \\ \hline
\end{tabular}
\end{table}

\begin{table}[]
\caption{Statistics of TPI on different $\varepsilon_{d}$} 
	\label{TrajStore_index_TPI_PI_varepsilon_d}  
\scriptsize
\begin{tabular}{ccccccccc}
\hline
\multicolumn{1}{l}{} & \multicolumn{2}{c}{Index Size(MB)} & \multicolumn{2}{c}{Time Cost} & \multicolumn{2}{c}{No.Periods} & \multicolumn{2}{c}{No.Insertions} \\ 
$\varepsilon_{d}$            & Porto           & Geolife          & Porto        & Geolife        & Porto         & Geolife        & Porto          & Geolife          \\ \hline
0.2                  & 862.0             & 249.2          & 1252           & 6535         & 1136           & 13958         & \textbf{4457}            & \textbf{55951}           \\ 
0.4                  & 860.0             & 238.2          & 497           & 4445          & 625            & 7953          & 5716            & 66400           \\ 
0.6                  & 859.9             & 236.5          & 480           & 3145          & 355            & 5670          & 6613            & 88033           \\ 
0.8                  & \textbf{857.4}             & \textbf{235.1}          & \textbf{465}           & \textbf{2848}          & \textbf{245}            & \textbf{3567}          & 7326             & 90554          \\ \hline
\end{tabular}
\end{table}
   
 \subsubsection{Dynamic Data Organization}
In this section, we analyze the proposed partition-based index (PI) and temporal PI (TPI) on $\{T_i\}$ with different $\varepsilon_{c}$ and $\varepsilon_{d}$, in terms of building time cost, the number of partitioned time periods,``Insertion" and ``Re-build".

 

Table \ref{TrajStore_index_TPI_PI_varepsilon_c} reports that as $\varepsilon_{c}$ increases, the index size gradually decreases, since a higher $\varepsilon_{c}$ provides a higher tolerance of reusing the previous structure by performing "insertions''.  Similar results are  observed for $\varepsilon_{d}$ in Table \ref{TrajStore_index_TPI_PI_varepsilon_d}, i.e., a higher $\varepsilon_{d}$ allows a PI reused for more timestamps.

 \begin{figure}[tb]
 \hspace{-5mm}
 \subfloat[PPQ-A]
 { \begin{tikzpicture}
\begin{axis}[
font=\scriptsize,
x tick label style={/pgf/number format/1000 sep=},
every axis y label/.style=
{at={(ticklabel cs:0.5)},rotate=90,anchor=near ticklabel},
clip=false,
xlabel={$\varepsilon_{p}$},
      ylabel={Running time (s)},
ybar,
    bar width=2pt,
    enlargelimits=0.1,
    width=2.6cm,
height= 3cm,
    axis on top,ymax=30,
    symbolic x coords={0.01,0.02,0.03,0.04,0.05},
 legend image code/.code={
                \draw [/tikz/.cd,bar width=2pt,yshift=-0.1em,bar shift=0pt]
                plot coordinates {(0cm,0.5em)};
            },
 legend style={at={(1.00,.65)}, anchor=south east}, legend columns=2,
    ]
\addplot[color=blue, fill=blue] coordinates {(0.01,19.19) (0.02,17.09) (0.03,15.73) (0.04,15.39) (0.05,13.55)};
\legend{Porto, font=6t}
\end{axis}
\end{tikzpicture}
 \begin{tikzpicture}
\begin{axis}[
font=\scriptsize,
x tick label style={/pgf/number format/1000 sep=},
every axis y label/.style=
{at={(ticklabel cs:0.5)},rotate=90,anchor=near ticklabel},
clip=false,
xlabel={$\varepsilon_{p}$},
      ylabel={Running time (s)},
ybar,
    bar width=2pt,
    enlargelimits=0.1,
    width=2.7cm,
height= 3cm,
    axis on top,ymax=30,
    symbolic x coords={0.01,0.02,0.03,0.04,0.05},
 legend image code/.code={
                \draw [/tikz/.cd,bar width=2pt,yshift=-0.1em,bar shift=0pt]
                plot coordinates {(0cm,0.5em)};
            },
 legend style={at={(1.00,.65)}, anchor=south east}, legend columns=2,
    ]
\addplot[color=red, fill=red] coordinates {(0.01,28.31) (0.02,26.00) (0.03,25.66) (0.04,25.65) (0.05,25.31)};
\legend{Geolife, font=6t}
\end{axis}
\end{tikzpicture}}\label{partition_efficiency_PPQ_A}
 \subfloat[PPQ-S]
 { \begin{tikzpicture}
\begin{axis}[
font=\scriptsize,
x tick label style={
/pgf/number format/1000 sep=},
every axis y label/.style=
{at={(ticklabel cs:0.5)},rotate=90,anchor=near ticklabel},
clip=false,
xlabel={$\varepsilon_{p}$},
      ylabel={Running time (s)},
ybar,
    bar width=2pt,
    enlargelimits=0.1,
    width=2.6cm,
height= 3cm,
    axis on top,ymax=40,
    symbolic x coords={0.1,0.2,0.3,0.4,0.5},
 legend image code/.code={
                \draw [/tikz/.cd,bar width=2pt,yshift=-0.1em,bar shift=0pt]
                plot coordinates {(0cm,0.5em)};
            },
 legend style={at={(1.00,.65)}, anchor=south east}, legend columns=2,
    ]
\addplot[color=blue, fill=blue] coordinates {(0.1,22.94) (0.2,16.50) (0.3,15.30) (0.4,15.08) (0.5,14.53)};
\legend{Porto, font=6t}
\end{axis}
\end{tikzpicture}
 \begin{tikzpicture}
\begin{axis}[
font=\scriptsize,
x tick label style={
/pgf/number format/1000 sep=},
every axis y label/.style=
{at={(ticklabel cs:0.5)},rotate=90,anchor=near ticklabel},
clip=false,
xlabel={$\varepsilon_{p}$},
      ylabel={Running time (s)},
ybar,
    bar width=2pt,
    enlargelimits=0.1,
    width=2.7cm,
height= 3cm,
    axis on top,ymax=40,
    symbolic x coords={1,2,3,4,5},
 legend image code/.code={
                \draw [/tikz/.cd,bar width=2pt,yshift=-0.1em,bar shift=0pt]
                plot coordinates {(0cm,0.5em)};
            },
 legend style={at={(1.00,.65)}, anchor=south east}, legend columns=2,
    ]
\addplot[color=red, fill=red] coordinates {(1,29.96) (2,27.01) (3,26.69) (4,26.44) (5,26.20)};
\legend{Geolife, font=6t}
\end{axis}
\end{tikzpicture}}\label{partition_efficiency_PPQ_S}
 \centering
 \caption{Temporal partitioning running time against different $\varepsilon_{p}$}\label{partition_efficiency_with_different_threshold}
 \end{figure}
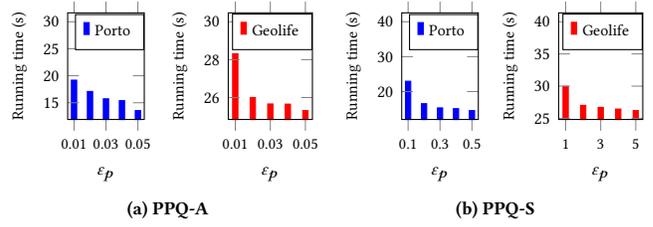
\begin{figure}[tb]
\centering  %
\subfloat[PPQ-A]{\label{partition_number_auto_porto}\includegraphics[height=2.3cm,width=4.5cm]{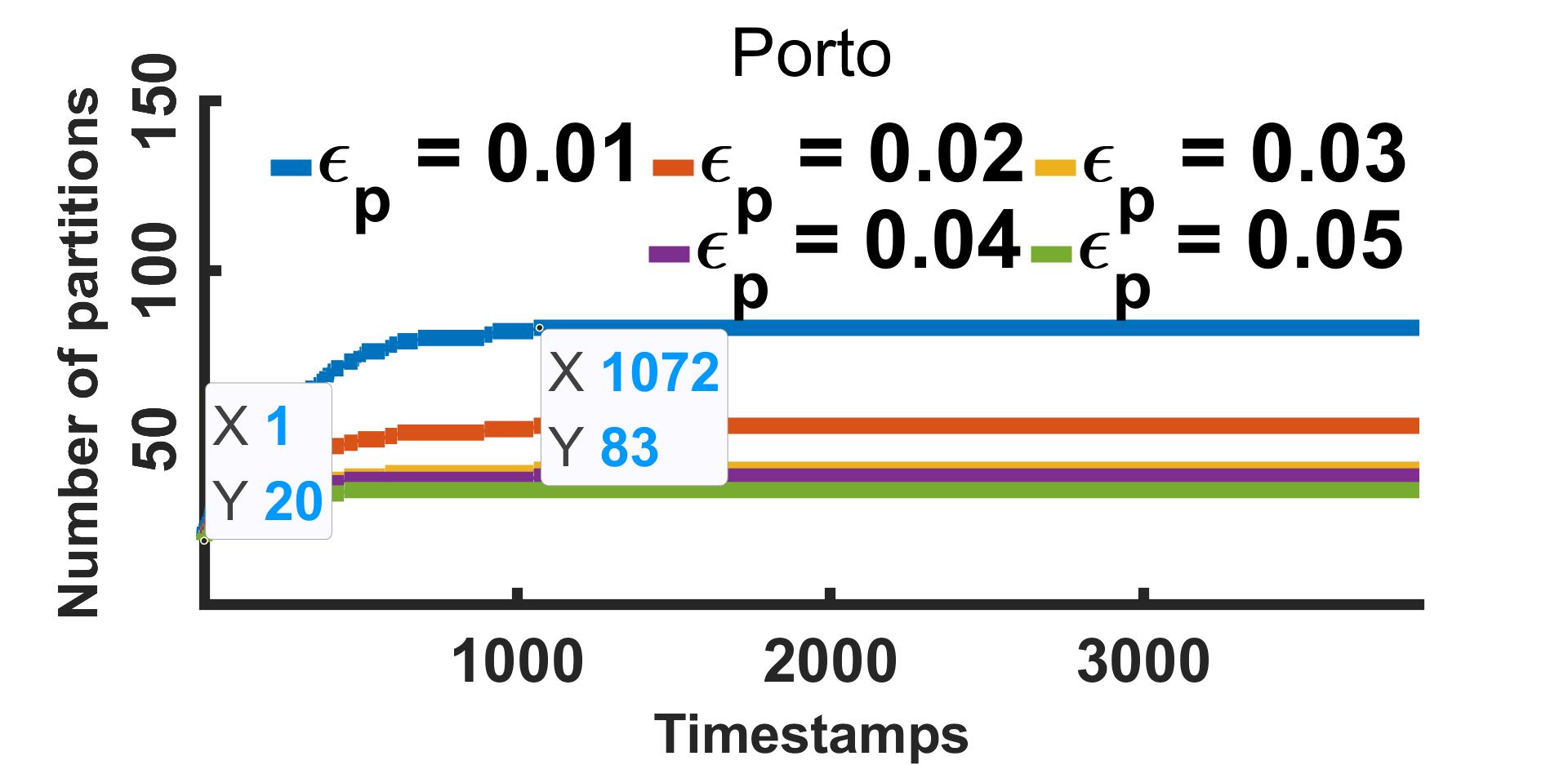}}
\subfloat[PPQ-A]{\label{partition_number_auto_geolife}\includegraphics[height=2.3cm,width=4.5cm]{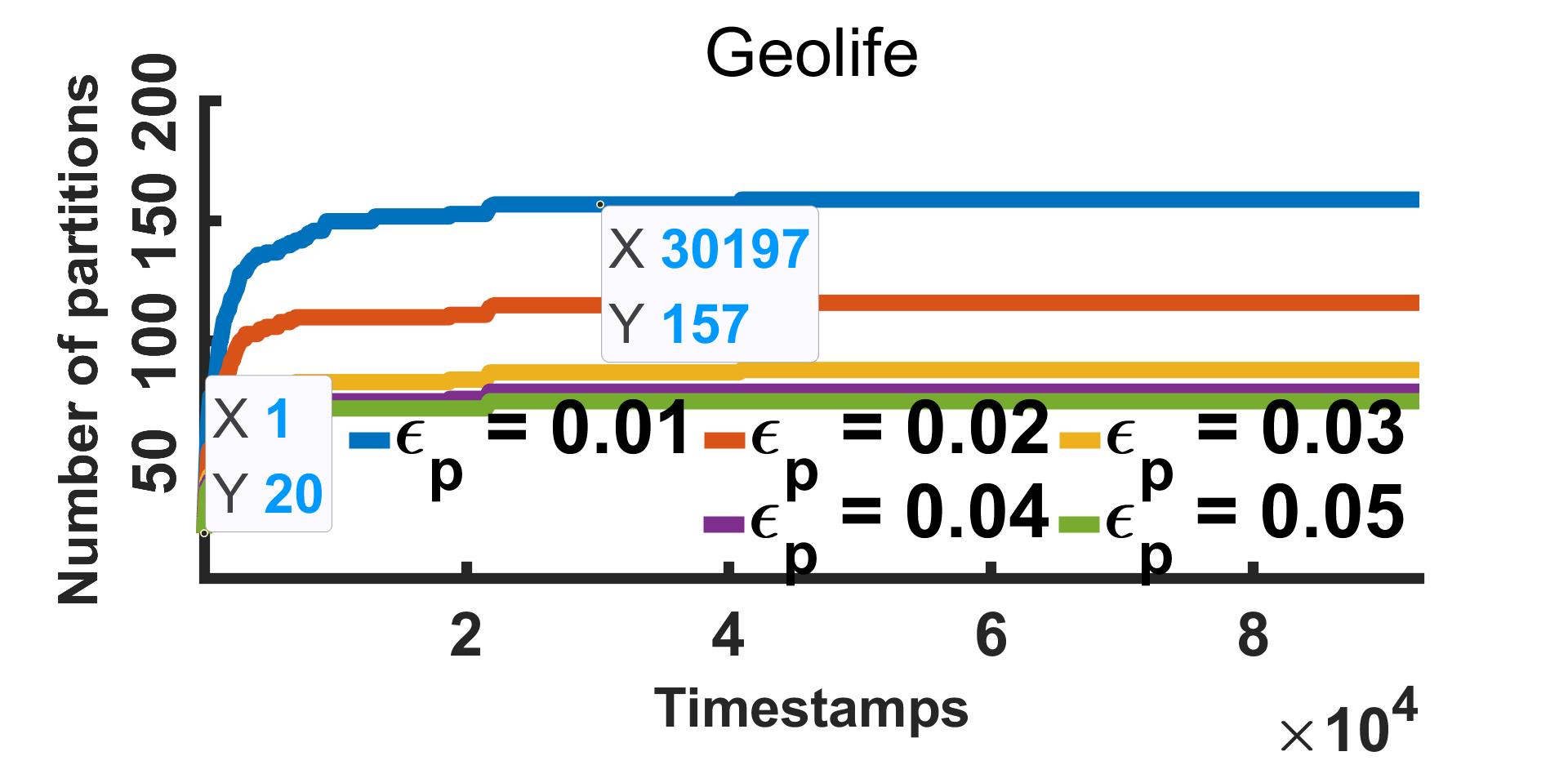}}

\subfloat[PPQ-S]{\label{partition_number_spatial_porto}\includegraphics[height=2.3cm,width=4.5cm]{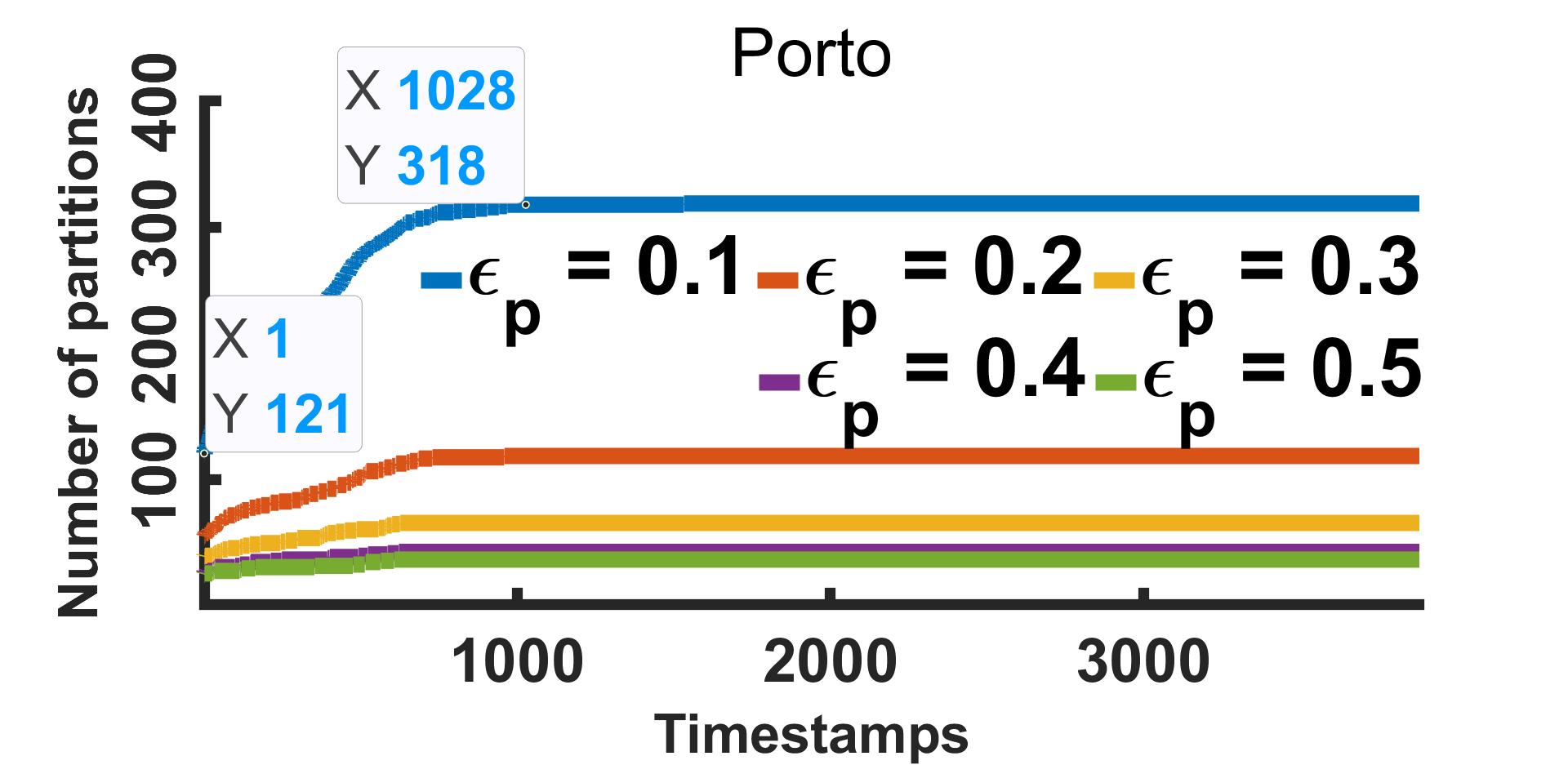}}
\subfloat[PPQ-S]{\label{partition_number_spatial_geolife}\includegraphics[height=2.3cm,width=4.5cm]{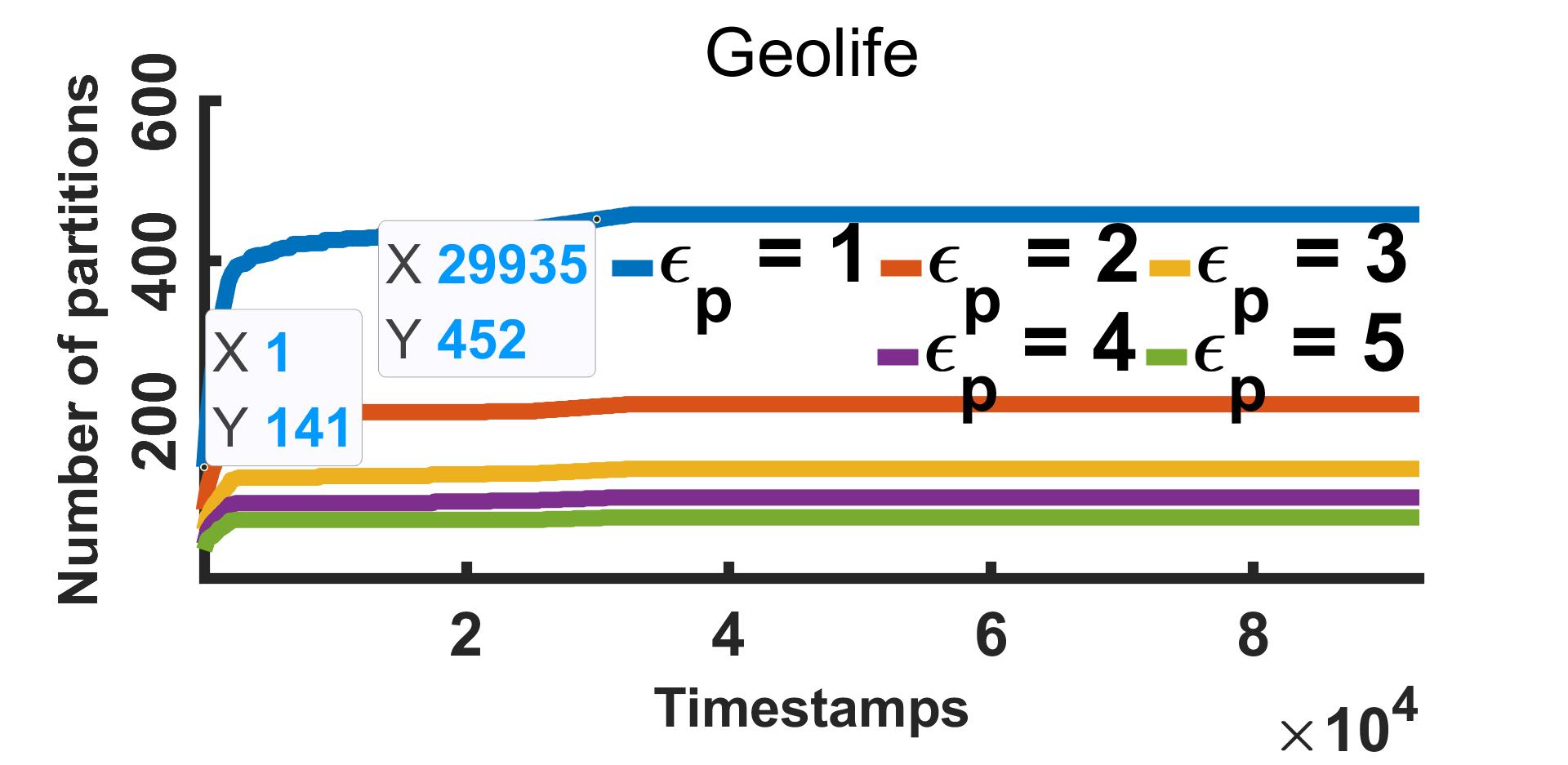}}
\caption{Number of partitions $q$ against different $\varepsilon_{p}$}\label{Number_of_partitions_Example}
\end{figure}
\subsubsection{Temporal Partitioning Efficiency}
In this section, we evaluate the efficiency of the incremental temporal partitioning (Section \ref{incremental_temporal_partitioning}), and analyze how the number of partitions change with time, with respect to different $\varepsilon_{p}$ values.

Figure \ref{partition_efficiency_with_different_threshold} illustrates that the running time of the temporal partitioning component reduces as $\varepsilon_p$ increases, since a smaller number of partitions is produced when $\varepsilon_p$ gets larger. In Figure \ref{Number_of_partitions_Example}, we also present the number of partitions that is maintained will gradually get stable with time increasing.
For example, in Figure \ref{partition_number_auto_porto},  we get the maximum number of partitions, 83, on Porto dataset at $t$ = 1072.
  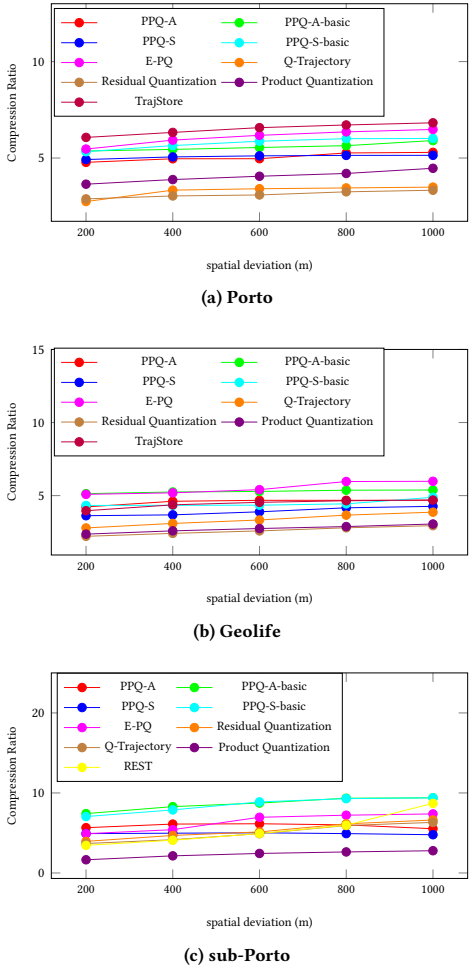
\begin{figure}[tb]
\subfloat[Porto]
 {\begin{tikzpicture}[scale=0.8]
\begin{axis}[
font=\scriptsize,
x tick label style={
/pgf/number format/1000 sep=},
every axis y label/.style=
{at={(ticklabel cs:0.5)},rotate=90,anchor=near ticklabel},%
xlabel={spatial deviation (m)},
      ylabel={Compression Ratio},
    width=8.5cm,
height= 5.2cm,
    ymax=13,
    symbolic x coords={200,400,600,800,1000},
 legend style={at={(.80,.481)}, anchor=south east}, legend columns=2,
    ]    
\addplot[color=red, mark=*] coordinates {(200,4.777) (400,4.962) (600,4.965) (800,5.263) (1000,5.293)}; 
\addplot[color=green, mark=*] coordinates {(200,5.365) (400,5.447) (600,5.546) (800,5.641) (1000,5.905)};
 \addplot[color=blue, mark=*] coordinates {(200,4.922) (400,5.058) (600,5.113) (800,5.137) (1000,5.140)};
\addplot[color=cyan, mark=*] coordinates {(200,5.329) (400,5.645) (600,5.872) (800,6.004) (1000,6.013)};
 \addplot[color=magenta, mark=*] coordinates {(200,5.465) (400,5.929) (600,6.177) (800,6.356) (1000,6.481)};
 \addplot[color=orange, mark=*] coordinates {(200,2.745) (400,3.335) (600,3.408) (800,3.450) (1000,3.492)};
 \addplot[color=brown, mark=*] coordinates {(200,2.882) (400,3.034) (600,3.086) (800,3.248) (1000,3.330)};
  \addplot[color=violet, mark=*] coordinates {(200,3.644) (400,3.883) (600,4.056) (800,4.200) (1000,4.468)};
  \addplot[color=purple, mark=*] coordinates {(200,6.069) (400,6.328) (600,6.575) (800,6.714) (1000,6.826)};
 
\legend{PPQ-A,PPQ-A-basic,PPQ-S,PPQ-S-basic,E-PQ, Q-Trajectory, Residual Quantization, Product Quantization, TrajStore}
\end{axis}
\end{tikzpicture}\label{porto_compression_ratio_against_deviation}}
 
 \centering
 \subfloat[Geolife]
 {\begin{tikzpicture}[scale=0.8]
\begin{axis}[
font=\scriptsize,
x tick label style={
/pgf/number format/1000 sep=},
every axis y label/.style=
{at={(ticklabel cs:0.5)},rotate=90,anchor=near ticklabel},%
xlabel={spatial deviation (m)},
      ylabel={Compression Ratio},
    width=8.5cm,
height= 5cm,
    ymax=15,
    symbolic x coords={200,400,600,800,1000},
 legend style={at={(.80,.48)}, anchor=south east}, legend columns=2,
    ]    
\addplot[color=red, mark=*] coordinates {(200,4.226) (400,4.612) (600,4.678) (800,4.678) (1000,4.687)}; 
\addplot[color=green, mark=*] coordinates {(200,5.133) (400,5.249) (600,5.285) (800,5.370) (1000,5.383)};
 \addplot[color=blue, mark=*] coordinates {(200,3.625) (400,3.683) (600,3.892) (800,4.165) (1000,4.265)};
\addplot[color=cyan, mark=*] coordinates {(200,4.328) (400,4.336) (600,4.345) (800,4.436) (1000,4.884)};
 \addplot[color=magenta, mark=*] coordinates {(200,5.087) (400,5.190) (600,5.408) (800,5.964) (1000,5.979)};
  \addplot[color=orange, mark=*] coordinates {(200,2.788) (400,3.097) (600,3.333) (800,3.665) (1000,3.867)};
 \addplot[color=brown, mark=*] coordinates {(200,2.216) (400,2.417) (600,2.586) (800,2.799) (1000,2.944)};
  \addplot[color=violet, mark=*] coordinates{(200,2.365) (400,2.589) (600,2.749) (800,2.882) (1000,3.057)};
  \addplot[color=purple, mark=*] coordinates{(200,3.966) (400,4.369) (600,4.540) (800,4.649) (1000,4.696)};
\legend{PPQ-A,PPQ-A-basic,PPQ-S,PPQ-S-basic,E-PQ,  Q-Trajectory, Residual Quantization,Product Quantization, TrajStore}
\end{axis}
\end{tikzpicture}\label{geolife_compression_ratio_against_deviation}}\\
 \centering

 \subfloat[sub-Porto]
 {\begin{tikzpicture}[scale=0.8]
\begin{axis}[
font=\scriptsize,
x tick label style={
/pgf/number format/1000 sep=},
every axis y label/.style=
{at={(ticklabel cs:0.5)},rotate=90,anchor=near ticklabel},%
xlabel={spatial deviation (m)},
      ylabel={Compression Ratio},
    width=8.5cm,
height= 5cm,
    ymax=25,
    symbolic x coords={200,400,600,800,1000},
 legend style={at={(.70,.47)}, anchor=south east}, legend columns=2,
    ]    
\addplot[color=red, mark=*] coordinates {(200,5.66) (400,6.11) (600,6.15) (800,6.01) (1000,5.53)}; 
\addplot[color=green, mark=*] coordinates {(200,7.42) (400,8.29) (600,8.74) (800,9.34) (1000,9.39)};
 \addplot[color=blue, mark=*] coordinates {(200,4.92) (400,4.97) (600,5.02) (800,4.92) (1000,4.78)};
\addplot[color=cyan, mark=*] coordinates {(200,7.05) (400,7.88) (600,8.88) (800,9.29) (1000,9.39)};
 \addplot[color=magenta, mark=*] coordinates {(200,4.91) (400,5.41) (600,6.96) (800,7.22) (1000,7.38)};
 \addplot[color=orange, mark=*] coordinates {(200,3.95) (400,4.69) (600,5.13) (800,6.14) (1000,6.62)};
 \addplot[color=brown, mark=*] coordinates {(200,3.68) (400,4.16) (600,4.91) (800,5.93) (1000,6.32)};
  \addplot[color=violet, mark=*] coordinates{(200,1.65) (400,2.14) (600,2.44) (800,2.63) (1000,2.78)};
  \addplot[color=yellow, mark=*] coordinates{(200,3.45) (400,4.09) (600,4.93) (800,5.97) (1000,8.70)};
\legend{PPQ-A,PPQ-A-basic,PPQ-S,PPQ-S-basic,E-PQ, Residual Quantization, Q-Trajectory, Product Quantization, REST}
\end{axis}
\end{tikzpicture}\label{rest_compression_ratio}}
 \centering
 \caption{Compression ratio against different
 spatial deviation}\label{compression_ratio_against_deviation}
 \end{figure}

\subsection{Compression Ratio}
In this section, the compression ratios of different methods are measured for different values of spatial deviation, following the same parameters setting as Section \ref{summary_efficiency}. The comparative results are presented in Figure \ref{compression_ratio_against_deviation}.
For the Porto dataset,  our solution outperforms Q-trajectory, residual quantization, and product quantization. The compression ratios of PPQ-A-basic and PPQ-S-basic slightly outperform  PPQ-A and PPQ-S, respectively, because PPQ-A and PPQ-S need additional space to store CQC.
The size of codebooks of  E-PQ and TrajStore is up to 11 and 27 times larger, respectively, than PPQ-trajectory (Table \ref{number_of_codewords}), however, they achieve compression ratios that are higher than PPQ-trajectory,
because we need additional space for multiple partitions, prediction coefficients $\{P_j[t]\}$ as well as CQC. Residual quantization and product quantization produce smaller sizes of codebooks compared to TrajStore, however, their compression ratios are 35\%--51\% smaller than that of TrajStore, because they need more space to store additional codeword indexes for restoring trajectory points from their summary. 
For Geolife dataset, PPQ-A-basic outperforms most of the alternatives in terms of the compression ratio, including Q-trajectory, residual quantization, product quantization and TrajStore. However, E-PQ produces compression ratios that are slightly higher than PPQ-trajectory when the spatial deviation is larger than 600m.
\begin{table}[t]
\caption{Disk-based index performance} 
	\label{TrajStore_index_TPI_PI}  
\scriptsize
\begin{tabular}{ccccccccc}
\hline
\multicolumn{1}{l}{} & \multicolumn{2}{c}{Index Size(MB)} & \multicolumn{2}{c}{No.I/Os} & \multicolumn{2}{c}{Response Time(s)} & \multicolumn{2}{c}{Building Time(s)} \\ 
Dataset              & Porto           & Geolife          & Porto       & Geolife       & Porto            & Geolife           & Porto            & Geolife           \\ \hline
TPI                  & \textbf{857.4}             & 235.1          & 1225           & 2230        & 24                 & 285             & \textbf{465}               & \textbf{2848}             \\ 
PI                   & 870.5             & 271.9          & \textbf{338}           & \textbf{301}        & \textbf{18}                 & \textbf{121}             & 1572              & 32009            \\ 
TrajStore            & \textbf{857.4}             & \textbf{233.5}          & 13803           & 35233        & 147                 & 378             & 4244              & 24372           \\ \hline
\end{tabular}
\end{table}


As mentioned in the experimental setting, REST have certain assumptions that do not hold for the general case we focus in this work. Hence, we only investigate the compression ratio on the sub-Porto dataset that is suitable for REST.  The compression ratios for different spatial deviations are shown in Figure \ref{rest_compression_ratio}.
The comparative results with respect to different spatial deviations are presented in Figure \ref{rest_compression_ratio}. When the spatial deviation is 200--600m, the compression ratios of PPQ-A-basic and PPQ-S-basic are two times that of REST. The gap decreases as the spatial deviation increases. REST's compression ratio depends on the correlation between the compressed trajectory and the reference set, the compressed trajectory cannot always be matched well with the offline learned reference set, which directly influences the compression ratio. However, PPQ-trajectory is able to flexibly extend codewords when the compressed trajectory can not be matched well with the existing codebook.
\subsection{Further Comparison with TrajStore}
In this section,  we provide disk-based comparisons of temporal partition-based index (TPI) with TrajStore in terms of the index size, query response times, the number of I/Os during queries, and building times. The index of TrajStore is built on the raw trajectory points, for fairness, we align disk-based TPI with TrajStore to directly build index over the raw trajectory points in accordance with the end of Section \ref{time_series_partition_based_index}.
We followed the same process in the TrajStore\cite{cudre2010trajstore}, bounding the data on disk and setting the page size as 1MB. We randomly select 10,000 spatio-temporal queries and sort them in the order of their starting times.  The parameters for TPI are $\varepsilon_{d}$=0.8 and $\varepsilon_{c}$=0.5. 

These experimental results are presented in Table \ref{TrajStore_index_TPI_PI}. The index size and building time of PI are larger than that of TPI and TrajStore for both datasets, while its response time and I/Os are the smallest among the three methods on both datasets.
For Porto, the size of TPI is the same as TrajStore, while the index size of TrajStore for Geolife is slightly smaller than that of TPI. 
However, TPI continuously outperforms TrajStore in terms of the number of I/Os and query response times. 
Given a query, TPI can quickly target the relevant $period_j$ and filter out more irrelevant trajectory points in terms of time.  However, for TrajStore, the quadtree-based index structure is shared by all timestamps, and a spatial cell can include trajectory points of a large time range, which might be stored on different pages. Hence, given a spatio-temporal query, for TrajStore several pages are likely to be visited, this is why the number of I/Os for TrajStore is over the number of queries we evaluate on. 

Building of TPI is  more efficient than TrajStore. 
TrajStore recursively updates the spatial index by merging, splitting or appending with trajectory points updated. However, when we get updates for TPI, it is only relevant to the trajectory points within a smaller range of timestamps, i.e., a $period_j$.

\section{Conclusions}
\label{conclusion}
We presented PPQ-trajectory which generates and maintains an error-bounded summary for large-scale trajectory data analytics. A partition-wise predictive quantizer (PPQ) for spatio-temporal data is designed, which involves a spatial proximity and autocorrelation based partitioning, followed by a local coding. A temporally-quantized data organization is developed to process spatio-temporal queries efficiently. The query performances, building times, and compression capabilities of PPQ-trajectory significantly outperform other solutions in most of the experiments. As a future work, the quantization based approach can be enhanced to consider dynamic traffic conditions, and utilize machine learning to more accurately predict trajectory points and generate a more compact summary.


\bibliographystyle{ACM-Reference-Format}
\balance
\bibliography{main} 

\end{document}